\documentclass[%
onecolumn,
superscriptaddress,
 amsmath,amssymb,
 aps, prfluids,
]{revtex4-2}

\usepackage{graphicx}
\usepackage{dcolumn}
\usepackage{bm}

\begin{document}

\preprint{APS/123-QED}

\title{\textbf{Interscale energy transfer in a propeller-driven homogeneous and isotropic turbulence generation device} 
}%

\author{Luciano Hergenreder}
\affiliation{Departamento de Ingeniería Hidráulica y Ambiental, Pontificia Universidad Católica de Chile, Av. Vicu\~{n}a Mackenna 4860, 7820436 Santiago, Chile}

\author{Serge Simoëns}
\affiliation{Laboratoire de Mécanique des Fluides et d’Acoustique, UMR CNRS 5509, Université Claude Bernard Lyon 1, INSA de Lyon, Ecole Centrale de Lyon, France}

\author{Cristián Escauriaza}
\affiliation{Departamento de Ingeniería Hidráulica y Ambiental, Pontificia Universidad Católica de Chile, Av. Vicu\~{n}a Mackenna 4860, 7820436 Santiago, Chile}

\author{Wernher Brevis}
\email[]{wbrevis@uc.cl}
\affiliation{Departamento de Ingeniería Hidráulica y Ambiental, Pontificia Universidad Católica de Chile, Av. Vicu\~{n}a Mackenna 4860, 7820436 Santiago, Chile}

\date{\today}

\begin{abstract}
We investigate the interscale energy dynamics of nearly homogeneous isotropic turbulence (HIT) generated in a spherically forced, propeller-driven device. From a spatial perspective, by employing the Kármán-Howarth-Monin (KHM) framework, we analyze the scale-by-scale energy transfer without isotropy assumptions. In contrast to other forcing strategies, where regions with inverse and forward cascades often coexist in scale space, we show that spherical forcing produces a purely forward energy cascade across the entire scale space. While the energy distribution remains highly isotropic and orientation-averaged structure functions align with Kolmogorov’s two-thirds law, the energy flux exhibits a directional dependence, suggesting a higher sensitivity to residual forcing asymmetries. Regarding the temporal dynamics, the reduced mean flow generated by this device allows us to investigate the temporal variability in conditions where Taylor's frozen turbulence hypothesis does not apply. Our results provide experimental validation of Tennekes' Eulerian frequency spectrum, where intermediate scales evidence an inertial-convective range with statistics determined by the integral velocity scale and the dissipation rate. Finally, we adopt a global approach to investigate whether the main turbulent scales can be linked to external experimental parameters. We propose a framework that captures the propeller frequency dependence of the main turbulent variables obtained in our measurements and those of previously reported HIT chambers. These findings contribute to the understanding of how external forcing influences the complex interscale energy dynamics of turbulent flows from a spatial, temporal, and global perspective.
\end{abstract}

\maketitle

\section{Introduction}\label{sec:introduction}
Understanding the transfer of energy in turbulent flows is fundamental to predicting the behavior of numerous natural and industrial systems. Turbulence is a complex, nonlinear phenomenon involving a wide range of scales and lacking a general analytical solution. In this context, the Richardson-Kolmogorov energy cascade has played a central role. It provides a conceptual framework that has driven significant advances, from the development of modeling approaches like RANS (Reynolds-Averaged Navier-Stokes) and LES (Large Eddy Simulation) to the interpretation of diverse physical processes, including frictional losses, scalar mixing, and sediment transport dynamics \citep{gioia2006turbulent,ali2018impact}.

The preferred flow configuration for studying fundamental properties of turbulence has long been homogeneous isotropic turbulence (HIT). Its inherent symmetries make it the simplest turbulent flow to analyze from a mathematical perspective, making it an ideal starting point for investigating the underlying flow dynamics of more complex cases \citep{batchelor1953theory,monin2013statistical}. However, experimentally generating HIT remains challenging, and a wide range of forcing strategies has been explored with varying results. Among these, the canonical grid-generated turbulence, while historically central to turbulence research, exhibits a strong mean flow that complicates Lagrangian measurements. Furthermore, it introduces a preferential direction of decay, being thus inhomogeneous along that axis. As an alternative, oscillating-grid turbulence avoids a strong mean flow but introduces large-scale secondary circulations \citep{variano2004random} that compromise HIT and also present a direction of decay. Additionally, von Kármán devices generate swirling flows characterized by a small central region where the mean velocity is low but highly anisotropic \citep{ouellette2006small,knutsen_inter-scale_2020}. More recent approaches inject energy using arrays of randomized jets, producing flows that are homogeneous \citep{Variano2008,carter_generating_2016,esteban2019laboratory} but only isotropic for certain configurations \citep{Bellani2014}. Motivated by these limitations, a new class of devices employing spherically distributed propellers within a closed chamber has gained popularity \citep{ghazi_nezami_laboratory_2023}. In these configurations, propellers mounted at the vertices of polyhedral \citep[e.g.,][]{birouk_attempt_2003,Zimmermann2010,Dou2016a} or spherical chambers \citep{bradley_measurement_2019,verwey_dissipation_2022} and oriented toward the chamber center, generate a central region of considerable homogeneity and isotropy with a reduced mean flow, providing a unique opportunity to explore the interscale energy dynamics under nearly-ideal conditions.

According to the classical picture of the Richardson-Kolmogorov energy cascade, turbulence comprises a hierarchy of eddies. The energy cycle begins at the largest of these structures, where energy is injected by the forcing mechanism. These eddies subsequently break down, transferring their energy to progressively smaller scales until they become sufficiently small for viscous effects to dissipate the kinetic energy \citep{Pope2000a}. During this cascade from large to small scales, the influence of the forcing mechanism is progressively lost, and the transfer takes place at a constant rate equal to the mean kinetic energy dissipation rate $\varepsilon$. According to \citet{kolmogorov1991local}, this implies that within an intermediate range of scales, known as the inertial subrange, the flow statistics are uniquely determined by $\varepsilon$. This foundational concept has led to the celebrated two-thirds and four-fifths laws for structure functions, as well as the $-5/3$ law for energy spectra.

Despite the central role of the turbulent cascade in the description of turbulent flows, it provides an idealized picture of the complex process of interscale energy transfer. 
In recent decades, significant progress in understanding interscale energy dynamics has been achieved through analyses within the framework of the Kármán-Howarth-Monin (KHM) equation. Formulated without assumptions of isotropy or homogeneity, the KHM equation offers a level of generality that enables a more complete characterization of the mechanisms governing interscale energy transfer. Analyses of different flows within this framework have revealed several discrepancies with the classical energy cascade picture, particularly highlighting the influence of energy injection on the dynamics across scales. For instance, \citet{alvesportela2017turbulence} showed that, in the flow behind an obstacle, regions of forward and backward energy transfer can coexist in scale space. A similar behavior was observed in the near field of a fractal grid \citep{gomes2015energy} and in homogeneous but anisotropic flows generated by randomly actuated jets \citep{carter2018small}. Notably, this effect appears in regions where conventional scaling laws still hold, such as $-5/3$ energy spectra and orientation-averaged structure functions consistent with the two-thirds law. In contrast, investigating a von Kármán flow, \citet{knutsen_inter-scale_2020} reported a purely forward energy cascade throughout scale space, with no regions exhibiting inverse transfer. Nevertheless, they observed a substantial energy transfer between scales of the same size, together with a marked degree of anisotropy, as well as a locally significant magnitude of the energy transfer due to mean flow. The impact of forcing is further evidenced in the results of \citet{valente2015energy}, who studied the flow behind regular grids. They observed that second- and third-order structure functions become increasingly isotropic with distance from the grid, while energy transfer becomes progressively dominated by large-to-small scales. Among the reported configurations, the oscillating grid flow of \citet{lamriben2011direct} shows the closest agreement with the classical energy cascade picture, exhibiting an isotropic energy distribution and a transfer that is both isotropic and strictly downscale. Together, these results emphasize the role of forcing in shaping interscale energy dynamics, motivating the present investigation in a spherically forced HIT chamber.

Evaluation of the interscale energy transfer is generally performed by means of spatial statistics, either in real or Fourier space, as in the KHM structure function analysis. Quantities measured in time are commonly mapped to the spatial domain via Taylor's frozen turbulence hypothesis \citep{taylor1938spectrum}. This hypothesis assumes that, if the mean velocity $U$ is much larger than the turbulent velocity $u^{\prime}$, the temporal response at a fixed point is the consequence of the passage of an unchanging pattern of turbulent motion convected uniformly at velocity $U$. This results in a linear relation between space and time. When it holds, scaling laws derived in the spatial domain should be directly observable in temporal statistics. However, when the mean flow is weak, the phenomenology underlying Taylor's hypothesis is no longer valid. Under these conditions, \citet{tennekes1975eulerian} proposed that the measured time signal does not exhibit an inertial subrange where the statistics are solely determined by the dissipation rate, as large-scale structures also influence these intermediate scales. This hypothesis has primarily been evaluated experimentally in homogeneous yet anisotropic flows, and here we leverage the salient characteristics of the propeller-driven HIT chamber to investigate it under nearly isotropic conditions.

Motivated by these observations, this study characterizes the interscale energy transfer in a spherically forced HIT chamber. First, we employ the Kármán-Howarth-Monin framework to evaluate the extent to which the flows produced by this specific energy injection strategy exhibit the key features of the Richardson-Kolmogorov energy cascade. Second, we leverage the ability of the device to generate nearly homogeneous and isotropic turbulence with a reduced mean flow, a condition rarely achieved in laboratory experiments. This allows us to analyze frequency energy spectra under conditions where the Taylor frozen turbulence hypothesis does not hold, providing direct insight into the temporal dynamics of the energy cascade and complementing Kármán-Howarth-Monin analysis, which is based on spatial statistics. Finally, after examining the interscale dynamics in detail, we investigate whether the dissipation rate and other characteristic scales of the energy cascade can be linked to external experimental parameters. To this end, we analyze from a global perspective the flow statistics produced by our device and those reported in the literature generated by other devices with the same forcing. The paper is organized as follows: Section~\ref{sec:methods} describes the experimental setup and measurement techniques; Section~\ref{sec:flow_characterization} presents the flow characterization; Sections~\ref{sec:khh} to \ref{sec:scaling} examine the energy cascade by the Kármán-Howarth-Monin analysis, temporal energy spectra, and global analysis; and Section~\ref{sec:conclusions} summarizes the main findings and discusses their implications.

\begin{figure*}
\begin{center}
\includegraphics[width=\textwidth]{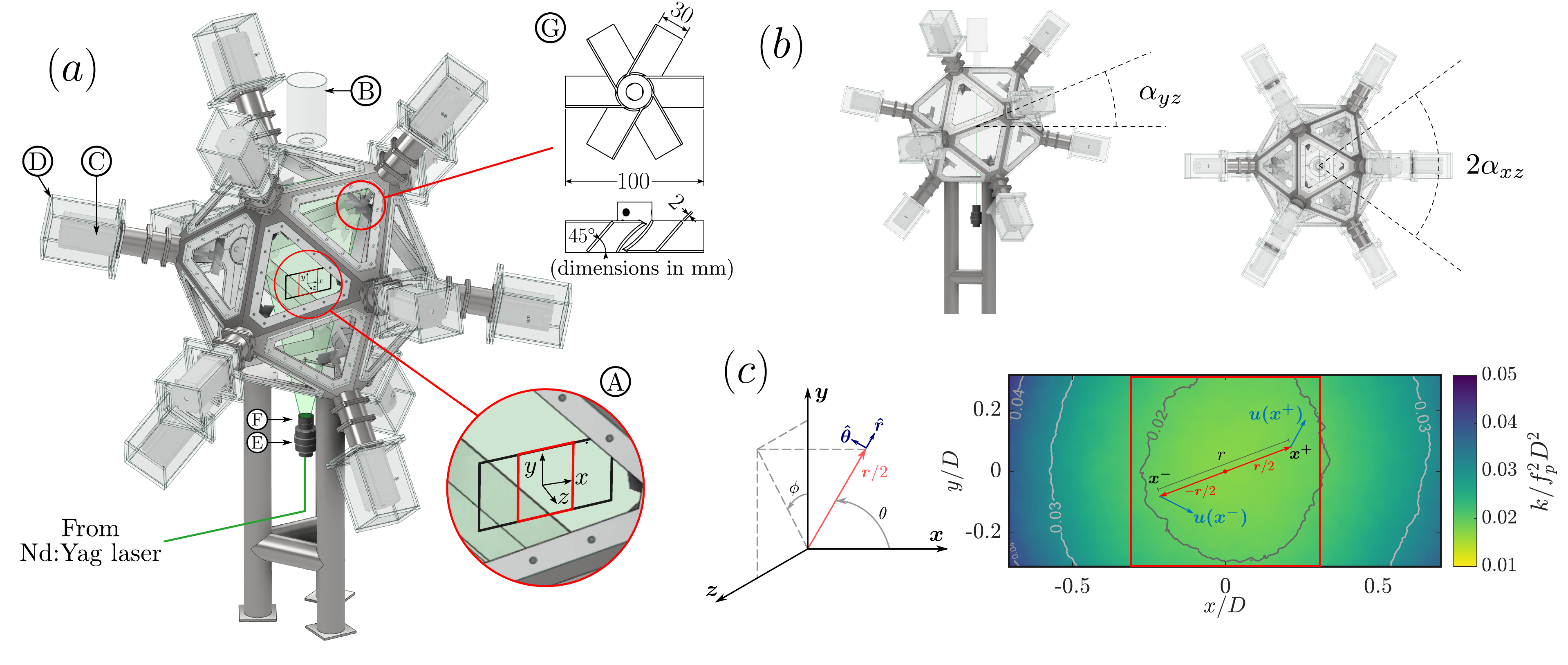}
\end{center}
\caption{\label{fig:setup} (a) Experimental setup. Field of view (black rectangle), central region (red square) and coordinate system (A); upper plate (B); motor (C); acrylic protection box (D); collimator (E); cylindrical lens (F); and propeller (G). (b) Camera viewing angles $\alpha _{xz}  \approx 37 ^{\circ}$ and $\alpha _{yz} \approx 23 ^{\circ}$. (c) Spherical coordinate system employed to compute two-point statistics, and representative TKE field, measured in the $xy$ ($\phi=0$) plane, where the central HIT region shown in red. The impact of the spherical forcing can be seen in that the isovalues are nearly circular and distributed around the center of the chamber.}
\end{figure*}

\section{Methods}
\label{sec:methods}

The experimental apparatus, based on the design of \citet{Zimmermann2010}, is a regular icosahedron with an edge length of 40 cm, enclosing 140 liters of water, used as the test fluid (Figure~\ref{fig:setup}). The flow is established by 12 propellers mounted at the vertices and oriented toward the center of the chamber. The propellers have a pitched blade geometry with a diameter $D=10$ cm and six blades of 30 mm width and 2 mm thickness (Figure~\ref{fig:setup}.(a)). They operate always in forward thrust at constant speeds $f_p $ reported in Table~\ref{tab:table1}. We characterize the flow conditions by means of a Reynolds number of the installation, $Re _I = f_p D^2/ \nu $ where $\nu$ is the kinematic viscosity of the fluid.

\begin{table}[b]
\caption{\label{tab:table1}
Summary of experimental parameters: the propeller rotational frequency $f_p$, the device Reynolds number $ Re _I = f_p D^2/\nu$, the r.m.s. velocity at the chamber center $\mathcal{U}=u^{\prime}_{ctr} = \langle \sqrt{2/3 k }\rangle _{ctr}$, the integral length scale $ L_{LL} $, the dissipation of turbulent kinetic energy per unit mass $\varepsilon = \langle \varepsilon_{LEPIV}\rangle_{ctr}$, the Kolmogorov length and time scales $\eta$ and $\tau _{\eta}$, the Taylor microscale $\lambda$ and the Reynolds number based on $\lambda$, $ Re _{\lambda} = u^{\prime}_{ctr} \lambda / \nu $; and the spatial resolution $\Delta x/\eta$.}
\begin{ruledtabular}
\begin{tabular}{ccccccccccc}
 $f_p$ [Hz]  & 
 $Re _I$ & $\mathcal{U}$ [mm/s] &  
 $L_{LL}$ [mm] & 
 $ \varepsilon$ [mm\textsuperscript{2}/s\textsuperscript{3}]
 & 
 $ \eta $ [mm] &
 $ \tau_{\eta} $ [ms] &
 $ \lambda $ [mm] & 
 $ Re _{\lambda} $ & 
 $ \Delta x / \eta $\\
\hline
3.00 & $2.7 \times 10^4$ & 33.75 & 42.62 & 369.76 & 0.247 & 55.82 & 7.30 & 225 & 2.8 \\
4.00 & $3.7 \times 10^4$ & 46.31 & 43.72 & 970.61 & 0.193 & 33.95 & 6.09 & 257 & 3.6 \\
5.02 & $4.6 \times 10^5$ & 57.92 & 44.44 & 1916.67 & 0.164 & 24.49 & 5.49 & 291 & 4.2 \\
5.99 & $5.5 \times 10^4$ & 70.57 & 42.42 & 3458.17 & 0.141 & 18.26 & 4.99 & 322 & 4.9 \\
7.17 & $6.6 \times 10^4$ & 84.59 & 45.52 & 5978.61 & 0.123 & 13.90 & 4.55 & 352 & 5.6 \\
\end{tabular}
\end{ruledtabular}
\end{table}

Stereo particle image velocimetry (SPIV) is employed to measure the three velocity components in a vertical plane centered in the chamber volume (Figure~\ref{fig:setup}~(a)). Two TSI PowerView 12MP cameras with Scheimpflug adapters and 85 mm Tokina lenses with apertures set to f\#11 are positioned such that their lines of sight are perpendicular to two of the windows. The resulting angle between cameras, measured in the $xz$ plane is $\alpha _{xz} \approx 37 ^{\circ}$. Additionally, a vertical angle (measured in the $yz$ plane) of $\alpha _{yz} \approx -23 ^{\circ} $ is present due to the geometry of the chamber (see Figure~\ref{fig:setup}~(b)). A dual-head 532~nm wavelength Nd:YAG laser (Beamtech Vlite-Hi-100) along with a -25~mm cylindrical lens and a collimator is used to generate the measurement plane from the bottom window of the chamber. 
The fluid is seeded with hollow glass particles with a density $\rho _p = 1.5 $ g/cm\textsuperscript{3}, and a mean diameter $d_p$ between 10-13 $\mu m$. The resulting Stokes number is $St = \tau_p / \tau_{\eta}$ below $ 10^{-6}$, ensuring that the particles behave as passive tracers \citep{adrian2011particle}. The particle images acquired by each camera are first processed using PIVlab \citep{thielicke_particle_2021}, by applying a contrast limited adaptive histogram equalization (CLAHE) and a low-pass filtering, followed by a two-pass PIV cross-correlation analysis using interrogation windows of $64 \times 64$ px and $32 \times 32$ px with 50\% overlap. Outlier vectors, identified using a local median criterion, account for less than 3\% of the total vector field. Finally, the resulting planar velocity fields for each camera are combined to obtain the two-dimensional, three component velocity field using an in-house implementation of the method of \citet{soloff_distortion_1997}. This results in a field of view (FOV) of $141 \times 65 $ mm with a vector separation of $\Delta x = 0.69$ mm (between 2.5 and 6 times the Kolmogorov scale, see Table~\ref{tab:table1}). For each experimental condition, 7250 image pairs are acquired in batches of 250 at a sampling frequency of 1 Hz, resulting in more than $3 \times 10^3$ statistically independent velocity fields.

\section{Flow characterization} \label{sec:flow_characterization}
\subsection{Homogeneity, isotropy and mean flow strength} \label{subsec:flow_quality}

\begin{figure}
\begin{center}
\includegraphics[width=.5\textwidth]{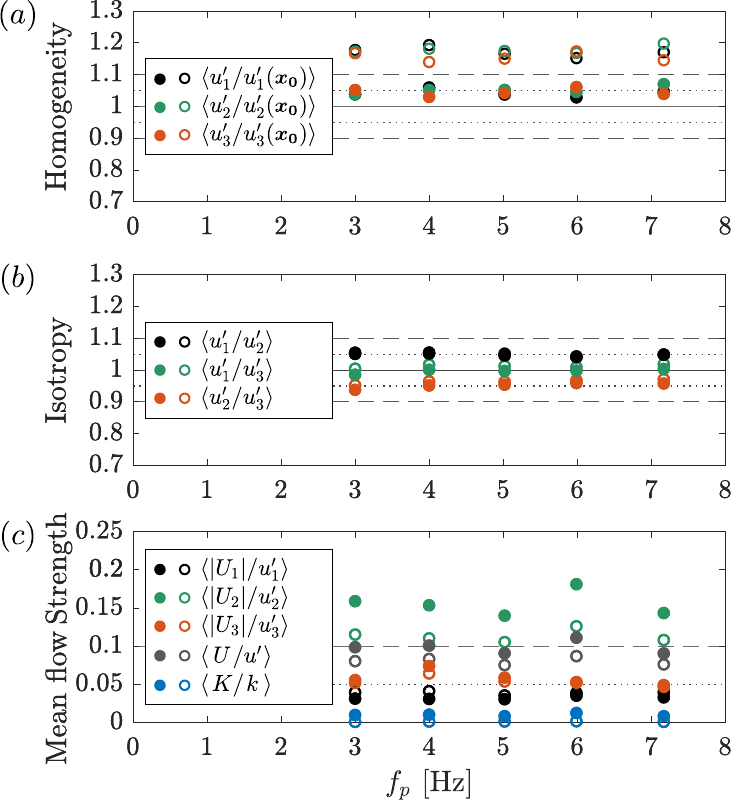}
\end{center}
\caption{\label{fig:flow-quality} Indices quantifying the degree of homogeneity (a), isotropy (b) and mean flow strength (c). Open and closed symbols correspond to spatial averages computed over the whole FOV and the central region shown in Figure~\ref{fig:setup}.(c), respectively.}
\end{figure}
\begin{figure*}
\begin{center}
\includegraphics[width=0.85\textwidth]{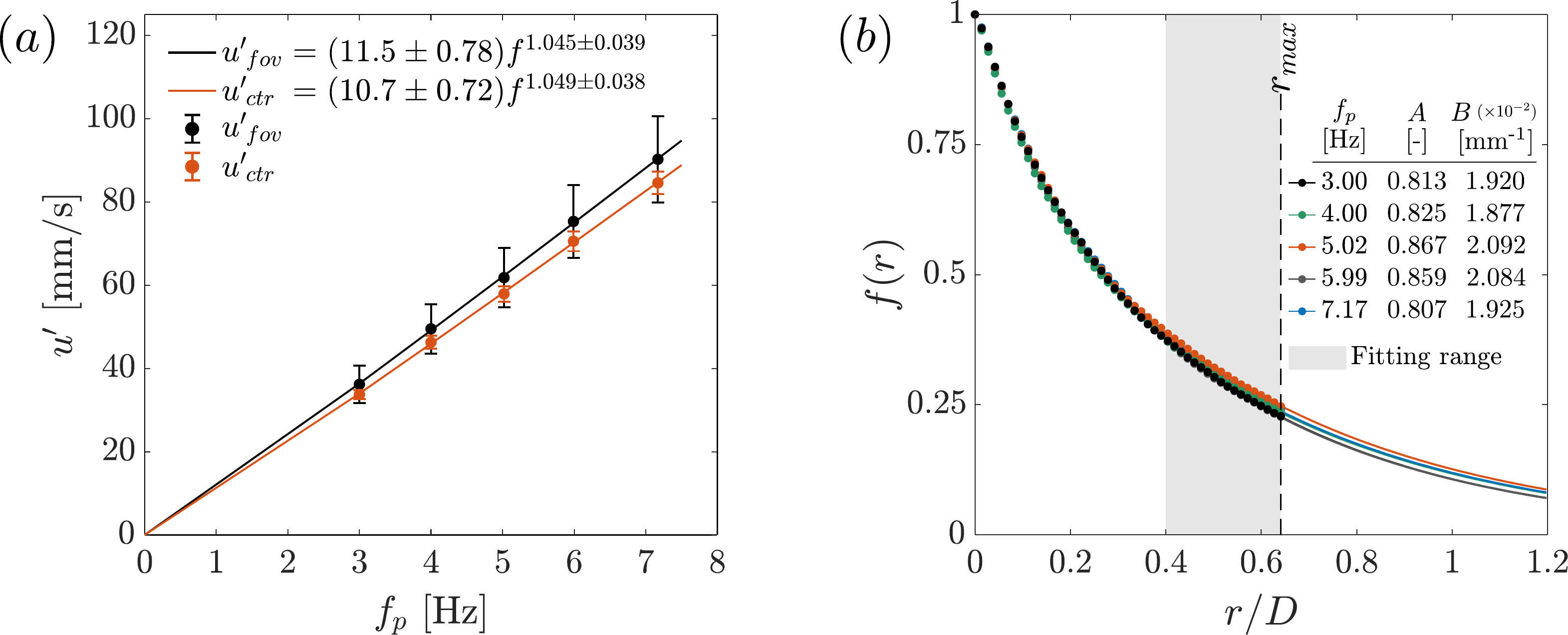}
\end{center}
\caption{\label{fig:length-velocity-scales} (a) Spatially averaged r.m.s. velocities as a function of propeller-frequency. Errorbars indicate twice the standard deviation obtained in space, and solid lines denote power-law fits. (b) Longitudinal autocorrelation functions for all forcing conditions (dots) with their exponential extrapolations (lines). The exponential tails were obtained by fitting the points within the shaded region, with the resulting fitting parameters $A$ and $B$ presented in the inset (see Eq.~\ref{eq:L_L}).}
\end{figure*}

The instantaneous velocity field obtained by SPIV, $V_i (\bm{x},t)$ $(i=1,2,3)$, is Reynolds decomposed into the mean $U_i = \overline{ V_i }$ and fluctuating parts $u_i(\bm{x},t)=V_i(\bm{x},t) - U_i(\bm{x}) $, where the operator $\overline{ (\cdot) }$ denotes time averaging. The turbulent and mean kinetic energy (TKE and MKE respectively) are subsequently defined as $k = \overline{ u_i u_i } / 2 $ and $K= U_iU_i /2$, while the root mean square (r.m.s.) of the $i$-th velocity component is given by $u_i ^{\prime} = (\overline{ u_i^2 })^{1/2}$. Spatial averages over regions $\Omega$ are denoted by brackets $\langle \cdot \rangle _{\Omega}$. In particular, we consider averages over the whole FOV, denoted as $\langle \cdot \rangle _{fov}$ and over the central region $ |x|/D \leq 0.3$ and $|y|/D \leq 0.3$, denoted as $\langle \cdot \rangle _{ctr}$ (Figure~\ref{fig:setup} (a) and (c)).

Figure~\ref{fig:flow-quality}~(a) presents a series of spatially averaged flow quality indices. The level of homogeneity is evaluated by considering the deviations of the fluctuating velocity components from its value at the chamber center, $\langle u^{\prime}_i(\bm{x})/u^{\prime}_i(\bm{x_0})\rangle$ \citep{chang_experimental_2012,Dou2016a}, where $\bm{x_0}=(0,0,0)$. Values averaged over the whole FOV present departures around $15\%$, while for the central region they fall within $\pm5\%$. The large-scale anisotropy, quantified by the spatial average of the ratio of r.m.s. velocity components, $\langle u^{\prime}_i/u^{\prime}_j \rangle$, is within approximately $5\%$ of the ideal value of unity, independent of the averaged area. The magnitude of the mean flow with respect to the fluctuating part, evaluated as $\langle |U_i|/u^{\prime}_i \rangle$ \citep{Variano2008}, is below $10\%$ except for the vertical component, which is slightly higher. However, other indices, such as the ratio of the mean to the turbulent kinetic energy, $\langle K/k\rangle$ \citep{Variano2008}, and the mean flow factor $\langle U/u^{\prime}\rangle = \langle \sqrt{2/3 K} / \sqrt{2/3 k} \rangle$ \citep{carter_generating_2016}, exhibit values below the thresholds of $5\%$ and $10\%$, respectively. Consequently, the mean flow is considered to be of negligible magnitude compared to the fluctuating part. Summarizing, the central region ($|x|/D , |y|/D\leq0.3$) presents levels of homogeneity, isotropy and mean flow strength comparable to all propeller-driven HIT devices \citep{ghazi_nezami_laboratory_2023}.

\subsection{Integral velocity and length scale} \label{sec:results_lengthscale}
We estimate the integral velocity scale $\mathcal{U}$, representative of the entire FOV and the central region by the spatially averaged r.m.s. velocities $u^{\prime}_{fov} = \langle \sqrt{2/3  k } \rangle _{fov} $ and $u^{\prime}_{ctr} = \langle \sqrt{2/3 k } \rangle _{ctr} $. Figure~\ref{fig:length-velocity-scales}~(a) illustrates the dependence of the integral velocity scale on the propeller frequency $f_p$. A power-law fit of the data reveals a linear dependence between both variables. The fact that the values corresponding to the entire FOV follow a linear trend, as do the data computed for the HIT region, shows that the scaling exponent is independent of the level of homogeneity of the flow. Since we concentrate on the central region for the rest of the paper, we adopt $\mathcal{U} = u^{\prime}_{ctr}$ hereafter and report these values in Table~\ref{tab:table1}.\\

To estimate the integral length scale $\mathcal{L}$, we analyze the spatial velocity correlation. For this, we follow \citet{lawson_unsteady_2022} and define two-point statistics around a central location $\bm{x}$. Relative to this central position, pairs of points separated by a distance $r = \lVert \bm{r} \rVert$ are defined as $ \bm{x^{-}} = \bm{x} - \bm{r} /2$ and $\bm{x^{+}} = \bm{x} + \bm{r} /2$. The velocity correlation tensor is then given by $R_{ij}(\bm{x},\bm{r})= \overline{ u_i( \bm{x^-},t) u_j(\bm{x^+},t) }$, and its projection along the direction of separation yields the longitudinal correlation $R_{LL}(\bm{x},\bm{r}) = R_{ij} \, r_i r_j /r^2$. To evaluate the velocity correlation as a function of the scalar separation $r = \lVert \bm{r} \rVert$, we adopt the spherical coordinate system illustrated in Fig.~\ref{fig:setup}~(c), defined by $(r,\theta,\phi)$. Here $\theta$ is the polar angle between $\bm{r}$ and $x_1$, and $\phi$ is the azimuthal angle, with $\phi=0$ corresponding to the measurement plane $xy$. Accordingly, the spherically averaged longitudinal correlation is given by
\begin{eqnarray}
\begin{split}
    \langle R_{LL} \rangle _{\circ} (\bm{x},r) 
    =& \frac{1}{4\pi} \int _{0} ^{2\pi} \int _{0} ^{\pi} R_{LL}(\bm{x},\bm{r}) \sin{\theta} d  \theta d \phi \\
    \approx& \frac{1}{2} \int _0 ^{\pi} \left. R_{LL} (\bm{x},\bm{r}) \right|_{\phi = 0} \sin \theta d\theta
\end{split}
\label{eq:saa}
\end{eqnarray}

\noindent where we assumed axisymmetry around $x_1$ \citep{lawson_unsteady_2022}. To determine the integral length scale, we first estimate the longitudinal autocorrelation function as $f(r)= \langle R_{LL}\rangle _\circ (\bm{x}_0,r)/ \langle R_{LL} \rangle_\circ (\bm{x}_0,r=0)$, where the spherically averaged correlation $\langle R_{LL} \rangle_\circ$ is evaluated at the geometric center of the chamber, $\bm{x}_0 = (0,0,0)$. The longitudinal integral length scale is then calculated via
\begin{eqnarray}
L_{LL} = \int _0 ^{\infty} f(r) \mathrm{d} r \approx \int _0 ^{r_{max}} f(r) \mathrm{d} r + \frac{A}{B} \exp (-B r_{max})
\label{eq:L_L}
\end{eqnarray}

\noindent where an exponential function is fitted to the tail of $f(r)$ to overcome the experimental limitation of measuring only finite separations \citep{DeJong2009}. Here, $r_{max}$ denotes the maximum separation resolved within the FOV, and $A$ and $B$ are the fitting parameters of the exponential function. The autocorrelation functions with their corresponding fits are presented in Figure~\ref{fig:length-velocity-scales}~(b) for all cases studied. The curves collapse for the different forcing conditions, which results in a fairly constant length scale across all cases. As detailed in Table~\ref{tab:table1}, the integral scales obtained for each propeller frequency exhibit a nearly constant value around 43.7 mm. The contribution of the extrapolation fit to $L_{LL}$ is approximately 25\% for all cases, which is the primary source of uncertainty limiting the precise determination of $L_{LL}$, as indicated by \citet{DeJong2009}.

\subsection{Dissipation rate} \label{sec:results_dissipation}
We perform an initial estimation of the dissipation rate based on velocity gradients to complement our analysis of Sec.~\ref{sec:khh}. Given that the spatial resolution of our velocity field is unable to resolve the Kolmogorov scale, $2.5 \leq \Delta x / \eta \leq 6$ (see Table~\ref{tab:table1}), we apply the large-eddy PIV method \citep{sheng2000large} to account for the smoothing of the velocity gradients due to this effect. This approach assumes dynamic equilibrium at a given scale $\Delta$, and estimates the dissipation rate by the energy flux below it. This implies $\varepsilon _{LEPIV}(\bm{x}) = -\overline{\tau_{ij} \tilde{s}_{ij}}$, where $\tilde{s}_{ij}=\frac{1}{2}(\partial \tilde{u}_i/\partial x_j + \partial \tilde{u}_j/\partial x_i)$ is the strain rate tensor based on the velocity filtered with a filter of size $\Delta$, and $\tau _{ij}$ the subfilter stress tensor. Considering the Smagorinsky model for $\tau _{ij}$, the dissipation rate can be expressed as:
\begin{equation}
    \varepsilon_{LEPIV}(\bm{x}) = 2 C_s ^2 \Delta ^2 \overline{(\tilde{s}_{ij}\tilde{s}_{ij})^{3/2}}
\label{eq:lepiv}
\end{equation}
\begin{figure}
\includegraphics[width=.85\textwidth]{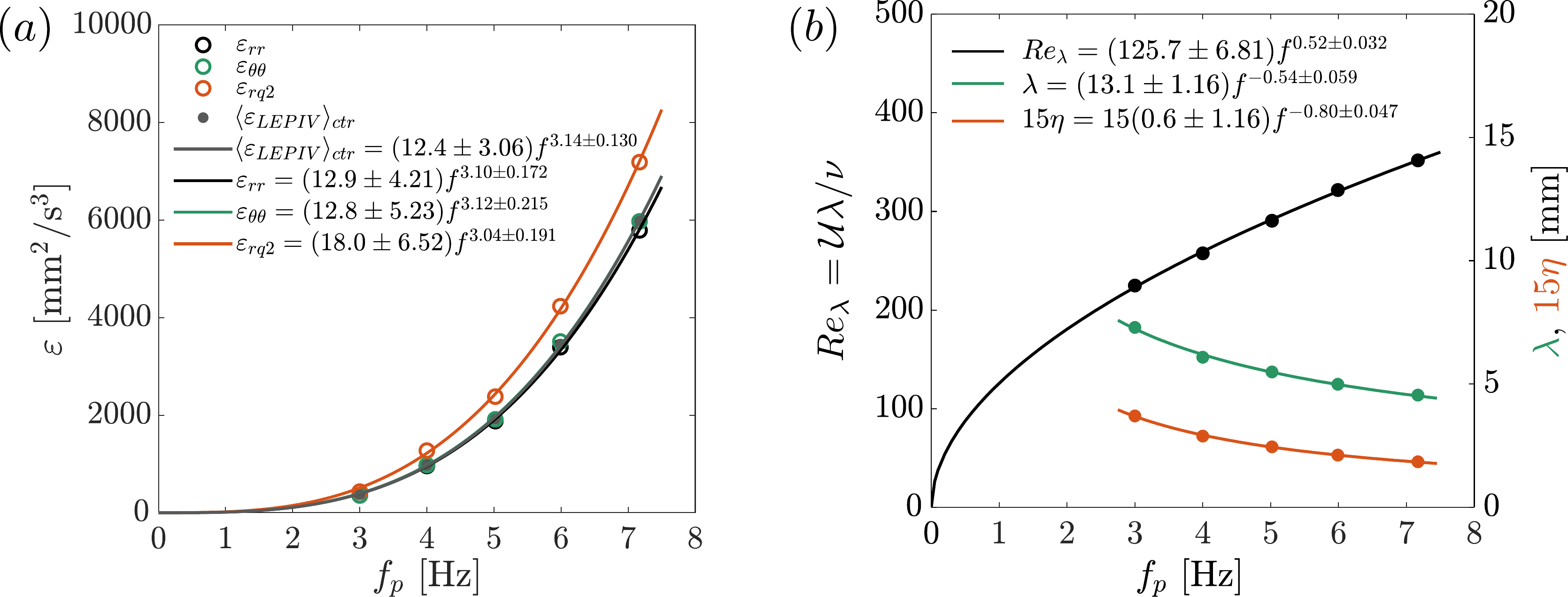}
\caption{\label{fig:dissipation_rates} (a) Dissipation rate as a function of propeller frequency, estimated via large-eddy PIV method averaged over the central region ($\langle \varepsilon_{LEPIV} \rangle_{ctr}$), and inertial-range predictions for the second-order longitudinal ($\varepsilon_{rr}$), second-order transverse ($\varepsilon_{\theta \theta}$), third-order ($\varepsilon_{rq^2}$) structure functions.  (b) Propeller frequency dependence of the Taylor and Kolmogorov microscales, and Taylor Reynolds number. Solid lines show the power-law fits to the experimental data.}
\end{figure}
\noindent where $C_s$ is the Smagorinsky constant. We apply a box filter and estimate the norm of the strain rate tensor $\tilde{s}_{ij}\tilde{s}_{ij}$ employing the surrogate expression introduced by \citet{sharp_piv_2001} (equation 10 in their work). We previously demonstrated that this formulation serves as a robust approximation of the unfiltered norm $s_{ij}s_{ij}$ (see Appendix of Ref.~\citep{hergenreder2026experimental}). Finally, we adopt a filter width $\Delta = 10.45\, \text{mm}$, and a Smagorinsky constant $C_s = 0.1$, in line with values employed in other propeller-driven HIT chambers \citep{DeJong2009,cheng2022effects}.

We apply this method to compute the dissipation rate over our measurement domain, and present the values averaged over the central region for the different propeller frequencies in Figure~\ref{fig:dissipation_rates}~(a). A least-squares fit to the data reveals that the dissipation rate exhibit an approximately cubic dependence on the propeller frequency. Throughout the rest of this paper we consider the dissipation rate obtained via the large-eddy PIV method averaged over the central region as the reference value, $\varepsilon = \langle \varepsilon_{LEPIV} \rangle_{ctr}$. We also use this reference value to estimate the Kolmogorov length and time scales, $\eta = (\nu^3/\varepsilon)^{1/4}$ and $\tau_\eta = (\nu/ \varepsilon)^{1/2}$, and the Taylor microscale $\lambda = (15 \nu \mathcal{U}^2/ \varepsilon)^{1/2}$ for each propeller frequency (Table~\ref{tab:table1} and Fig.~\ref{fig:dissipation_rates}~(b)).

\section{Spatial perspective: Kármán-Howarth-Monin analysis} \label{sec:khh}
Having presented the overall properties of the turbulent flows, this section focuses on interscale energy transfer using the Kármán-Howarth-Monin framework. The KHM (also referred to as Kármán-Howarth-Monin-Hill or Kármán-Howarth-Hill) equation describes the interscale energy dynamics through two-point statistics. Following Sec.~\ref{sec:results_lengthscale}, we consider two points located symmetrically around a central position $\bm{x}$, separated by the vector $\bm{r}$ with a magnitude $r = \lVert \bm{r} \rVert$. These points are defined as $ \bm{x^{-}} = \bm{x} - \bm{r} /2$ and $\bm{x^{+}} = \bm{x} + \bm{r} /2$. Accordingly, the differences of the fluctuating velocity, mean velocity and pressure are defined as, $\delta u_i \equiv u_i^+ - u_i ^-$, $\delta U_i \equiv U_i^+ -  U_i^-$ and $\delta p \equiv p^+ - p^-$, respectively, where the superscripts denote statistics at the corresponding spatial points.
Given the high degree of homogeneity in our flow, as discussed in Sec.~\ref{subsec:flow_quality}, we apply a spatial average over the central region to the two-point statistics, complementing the temporal averaging \citep{lamriben2011direct,knutsen_inter-scale_2020,wang2025energy}. Consequently, we base our analysis on the spatially averaged version of the KHM equation, which is expressed as \citep{knutsen_inter-scale_2020},

\begin{align}
\underbrace{\frac{\partial \langle \overline{\delta q^2} \rangle_{\Omega}}{\partial t}}_{4\mathcal{A}_t}
+ \underbrace{\left(\frac{U_k^{+} + U_k{^-}}{2}\right) \frac{\partial \langle  \overline{\delta q^2} \rangle_{\Omega}}{\partial x_k}}_{4\mathcal{A}} 
+ \underbrace{\frac{\partial \langle \overline{\delta u_k \delta q^2} \rangle_{\Omega}}{\partial r_k}}_{4\Pi} 
+ \underbrace{\frac{\partial \langle \delta U_k \overline{\delta q^2} \rangle_{\Omega}}{\partial r_k}}_{4\Pi_U} \nonumber \\
= \underbrace{-2 \left\langle \overline{\delta u_i \delta u_k}  \frac{\partial \delta U_i}{\partial r_k} \right\rangle_{\Omega}
- \left\langle \overline{ (u_k^{+} + u_k^{-}) \delta u_i}  \frac{\partial \delta U_i}{\partial x_k} \right\rangle_{\Omega}}_{4\mathcal{P}} 
\underbrace{-\frac{1}{2} \left\langle \frac{ \partial \overline{(u_k^{+} + u_k^{-}) \delta q^2} }{\partial x_k} \right\rangle_{\Omega} }_{4\mathcal{T}_u} \,
\underbrace{-\frac{2}{\rho} \left\langle \frac{\partial \overline{ \delta u_k \delta p}}{\partial x_k} \right\rangle_{\Omega}}_{4\mathcal{T}_p} \nonumber \\
+ \underbrace{\frac{\nu}{2} \left\langle \frac{\partial^2}{\partial x_k^2} \overline{ \delta q^2 } \right\rangle_{\Omega}}_{4\mathcal{D}_x}
+ \underbrace{2 \nu \left\langle \frac{\partial^2}{\partial r_k^2} \overline{ \delta q^2 } \right\rangle_{\Omega}}_{4\mathcal{D}_\nu} 
-\underbrace{\frac{\nu}{2} \left[ \left\langle \overline{ \left( \frac{\partial u_j^+}{\partial x_i^+} \right)^2 } \right\rangle_{\Omega} +
\left\langle \overline{ \left( \frac{\partial u_j^-}{\partial x_i^-} \right)^2  } \right\rangle_{\Omega} \right]}_{4\varepsilon^*} 
\label{eq:khh_averaged}
\end{align}

\noindent where $\Omega$ denotes the spatial averaging domain, defined here as the central HIT region ($|x/D|,|y/D| \leq 0.3$).

\begin{figure}
\includegraphics[width=.5\textwidth]{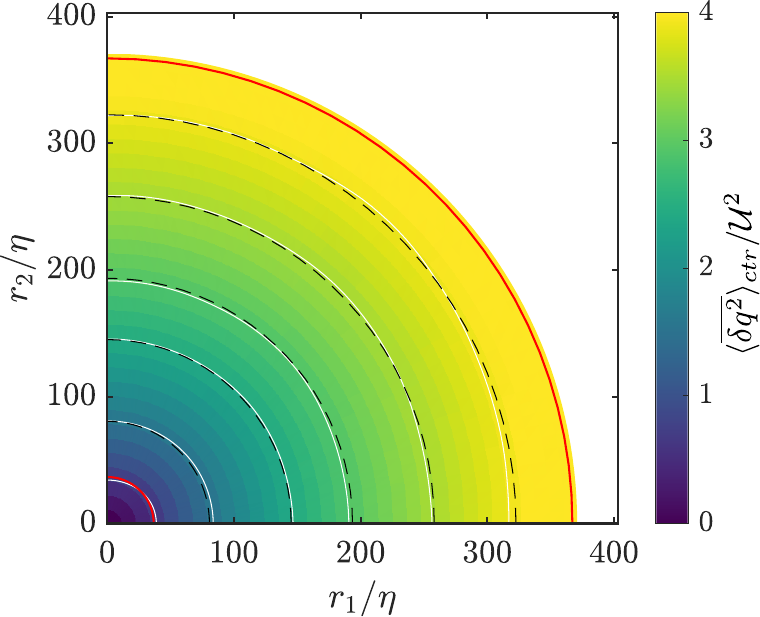}
\caption{\label{fig:khh_second_order_7.17Hz} Normalized second order structure function corresponding to the 7.17~Hz propeller frequency. White lines correspond to isocontours of $\langle \overline{\delta q^2}\rangle_{ctr}$, and dashed lines are circles included as reference of perfect isotropy. The innermost red circle denotes the size of the Taylor microscale, while the outermost corresponds to the integral length scale.}
\end{figure}

The expression~(\ref{eq:khh_averaged}) is the evolution equation of the second-order structure function $\overline{ \delta q^2} (\bm{r}) \equiv \overline{\delta u_i \delta u_i}$, which is conventionally considered as a measure of the fluctuating energy at scale $r = \lVert \bm{r} \rVert$ or less \citep{davidson2015turbulence}. As an energy budget, it accounts for the processes of production, transfer and dissipation in both physical $\bm{x}$ and scale space $\bm{r}$. On the left-hand side we have the spatially averaged unsteady term $\mathcal{A}_t$, advection $\mathcal{A}$, and the nonlinear ($\Pi$) and linear ($\Pi_U$) interscale transfer rates. The first two terms in the right-hand side describe the spatially averaged production of $\overline{\delta q^2}$, denoted by $\mathcal{P}$. They are followed by terms capturing the spatially averaged transport of $\overline{\delta q^2}$ by the fluctuating velocity in physical space, $\mathcal{T}_u$; the spatially averaged pressure-velocity term, $\mathcal{T}_p$; the spatially averaged viscous diffusion in physical $\mathcal{D}_x$ and scale space $\mathcal{D}_\nu$; and the spatially averaged two-point dissipation $\varepsilon^*$. For further details on each term the reader is referred to Refs.~\citep{valente2015energy,gomes2015energy,knutsen_inter-scale_2020}.

According to \citet{vassilicos2015dissipation} and \citet{gomes2015energy}, if a turbulent flow at high Reynolds numbers is locally homogeneous for a sufficiently small scale $r<\mathcal{L}$, the linear energy transfer $\Pi_U$ and all terms on the right-hand side of Eq.~(\ref{eq:khh_averaged}) vanish, except for $\mathcal{D}_\nu$ and $\varepsilon^*$. Furthermore, under these conditions, the two-point dissipation rate $\varepsilon^*$ reduces to $\varepsilon$. Additionally, at high Reynolds numbers, the viscous diffusion $\mathcal{D}_\nu$ can be neglected for scales larger than the Taylor microscale. This leaves only the unsteady term ($\mathcal{A}_t$), advection term ($\mathcal{A}$), nonlinear energy transfer ($\Pi$) and dissipation ($\varepsilon$), which govern the energy budget across the locally homogeneous scale range $\lambda < r < \mathcal{L}$. Next, the local equilibrium hypothesis of \citet{kolmogorov1991local} states that small-scale motions evolve very quickly compared to the overall turbulence evolution and are thus in statistical equilibrium, $\mathcal{A}_t + \mathcal{A} \approx  0$. This ultimately leaves a balance between nonlinear energy transfer and dissipation,
\begin{align}
    \Pi  \approx -\varepsilon  \label{eq:kolmogorov_balance}
\end{align}
\noindent where the nonlinear energy transfer is the divergence of the third order structure function in scale space, $\Pi = (1/4)\,\partial \langle \overline{\delta u_k \delta q^2} \rangle_\Omega/\partial r_k$. This balance is at the core of the dimensional analysis leading to the celebrated $2/3$ and $-5/3$ laws. Furthermore, adding the hypothesis of small scale isotropy of the third order structure function $\overline{\delta u_i \delta q^2}$ is also considered, Eq.~(\ref{eq:kolmogorov_balance}) can be reduced to \citep{gomes2015energy},
\begin{align}
    \langle \overline{\delta u_r \delta q^2} \rangle_\Omega \approx -\frac{4}{3} \varepsilon r \label{eq:third_order_durdq2}
\end{align}
\noindent where $\langle \overline{\delta u_r \delta q^2} \rangle_\Omega =\langle \overline{\delta u_i \delta q^2} \rangle_\Omega \cdot \hat{r}_i$ is the projection of the third order structure function in the radial direction. Equation~(\ref{eq:third_order_durdq2}) is a generalization of the four-fifths law, originally derived for longitudinal structure functions \citep{antonia1997analogy}.

\subsection{Energy transfers in scale space}
In this work we focus on the second-order $\langle \overline{\delta q^2} \rangle_{ctr}$ and third-order structure functions $\langle \overline{\delta u_i \delta q^2} \rangle_{ctr}$, as well as the linear and nonlinear energy transfer terms $\Pi$ and $\Pi_U$. The nonlinear interscale energy transfer, $\Pi = (1/4) (\partial \langle \overline{ \delta u_k \delta q^2 }\rangle_{ctr}/\partial r_k)$, accounts for the redistribution of energy $\overline{\delta q^2}$ in scale space due to nonlinear interactions. It is the divergence, in scale space, of the flux given by the third order structure function $\langle \overline{\delta u_k \delta q^2} \rangle_{ctr}$. Similarly, the linear energy transfer, $\Pi_U = (1/4) (\partial \langle \delta U_k \overline{ \delta q^2 }\rangle_{ctr}/\partial r_k)$, accounts for the interscale transfer of $\overline{\delta q^2}$ by effect of the mean flow.

We analyze the results in scale space considering the spherical coordinate system $(r,\theta,\phi=0)$ where $\theta$ is the polar angle between $\bm{r}$ and the zenith direction (defined by $x_1$), and $\phi$ is the azimuthal angle. Here $\phi=0$ corresponds to the measurement plane $xy$ over which the statistics are computed (Fig.~\ref{fig:setup}~c). Relative to the $r_1 - r_2$ axes, the polar angle $\theta$ is measured from the horizontal separation axis $r_1$ (aligned with the spatial coordinate $x_1$) toward $r_2$ (aligned with $x_2$), such that $\theta = 0$ corresponds to separations along $r_1$ and $\theta = \pi/2$ along $r_2$. The statistics are initially calculated in the Cartesian coordinates defined by the PIV grid and subsequently interpolated onto a polar grid. This polar grid is defined with a radial resolution matching the space between vectors, and a discretization of 56 angles between $0$ and $\pi/2$.

\begin{figure*}
\begin{center}
\includegraphics[width=\textwidth]{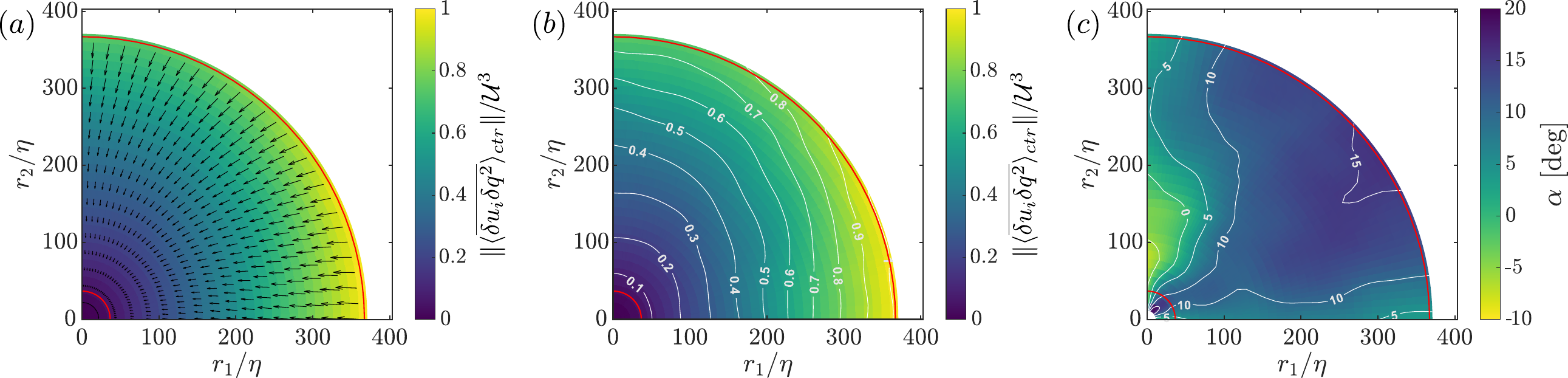}
\end{center}
\caption{\label{fig:water_khh_third_order_7.17Hz} Normalized third order structure function corresponding to the 7.17~Hz propeller frequency (a), isocontours of its magnitude (b), and angle between the vector field and the radial direction, $\alpha$ in degrees (c). The innermost red circle denotes the size of the Taylor microscale, while the outermost corresponds to the integral length scale. Only half of the vectors are plotted in (a).}
\end{figure*}

Figure~\ref{fig:khh_second_order_7.17Hz} presents the second-order structure function obtained for a representative case. We observe that its distribution throughout the entire scale space is almost perfectly isotropic (i.e., independent of the orientation considered), as denoted by the circular shape of its isocontours. This indicates an even energy distribution across scales of the same size, which is in agreement with the results of \citet{Zimmermann2010} who observed the independence of second-order longitudinal structure functions with the orientation of the separation vector $\bm{r}$ for a propeller-driven HIT chamber with the same geometry as ours. Figure~\ref{fig:water_khh_third_order_7.17Hz}~(a) presents the third-order structure function, where the vectors indicate the direction of the energy flux and the colors its magnitude. First, we observe that the magnitude displays an anisotropic distribution in scale space, with larger values observed around $\theta=0^\circ$ (corresponding to the horizontal separation axis $r_1$), resulting in approximately elliptic isocontours (Fig.~\ref{fig:water_khh_third_order_7.17Hz}~b). Regarding the direction of the energy flux, the vectors point mainly radially inward throughout the entire scale space. This is in line with an energy flux directed primarily from large to small scales, as conceptualized in the Richardson-Kolmogorov energy cascade. We examine this more closely by quantifying $\alpha$, the angle  between $\langle \overline{\delta u_i \delta q^2} \rangle_{ctr}$ and the inward radial direction $-\bm{r}$ (Fig.~\ref{fig:water_khh_third_order_7.17Hz}~c). We observe the largest deviations occur around $\theta \approx 45^\circ$, with maximum values of about $15^\circ$. These results indicate that the energy flux is predominantly from large to small scales, with minor transfer in the polar direction, which corresponds to redistribution of $\overline{\delta q^2}$ within scales of the same size.

As discussed by \citet{alvesportela2017turbulence}, while the fact that $\langle \overline{\delta u_i \delta q^2}\rangle_{ctr}$ points in an inward direction indicates a flux from large to small scales, this is not sufficient to determine whether the energy cascade is direct or inverse. A direct energy cascade requires inward fluxes locally coexisting with negative values of the radial component of the nonlinear interscale transfer $\Pi_r$,

\begin{align}
    \Pi_r  = \frac{1}{4} \frac{1}{r^2} \frac{\partial}{\partial r} \left( r^2 \langle \overline{\delta u_r \delta q^2} \rangle_{ctr} \right) \label{eq:Pi_r}
\end{align}

\begin{figure}
\includegraphics[width=.5\textwidth]{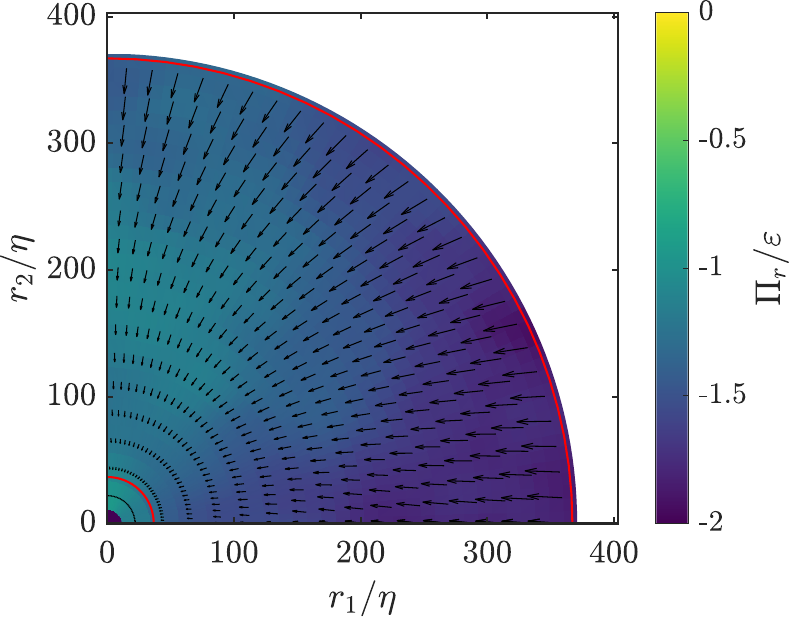}
\caption{\label{fig:khh_Pi_r_7.17Hz} Normalized radial component of the nonlinear energy transfer corresponding to the 7.17~Hz propeller frequency. The innermost red circle denotes the size of the Taylor microscale, while the outermost corresponds to the integral length scale. Only half of the vectors are plotted.} 
\end{figure}

To evaluate whether our flow exhibits a direct energy cascade, we present $\Pi_r$ in Figure~\ref{fig:khh_Pi_r_7.17Hz} and observe that it remains negative throughout the entire scale space. This negative behavior, combined with the inward direction of $\langle \overline{ \delta u_i \delta q^2} \rangle_{ctr}$, allows us to conclude that a direct energy cascade, from large to small scales, takes place across the entire scale space and that no inverse cascade exists. In addition, we note that the magnitude of $ \Pi_r$ varies with the orientation considered, being larger for $\theta\approx 0^\circ$, as expected by the magnitude of the flux presented in Fig.~\ref{fig:water_khh_third_order_7.17Hz}~(b).

We now consider the linear interscale energy transfer $\Pi_U$. Analogous to the nonlinear transfer term $\Pi$, it represents the redistribution of energy across scales, but here it is associated with the flux $\delta U_i \overline{\delta q^2}$, where $\delta U_i$ is the two-point difference of the mean velocity field. Under homogeneous conditions, with a weak mean flow, its contribution to the interscale energy balance should be negligible. However, \citet{knutsen_inter-scale_2020} analyzed a von Kármán flow in a low-mean-flow region, where $U_i/u_i^\prime < 5\%$ (no index summation), and reported a significant contribution of $\Pi_U$. Although the the orientation-averaged contribution of $\Pi_U$ was modest, both its magnitude and the magnitude of the linear flux $\delta U_i \overline{\delta q^2}$ were observed to be of the order of the nonlinear transfer $\Pi$ and nonlinear flux $\overline{\delta u_i \delta q^2}$, respectively.

\begin{figure}
\includegraphics[width=.85\textwidth]{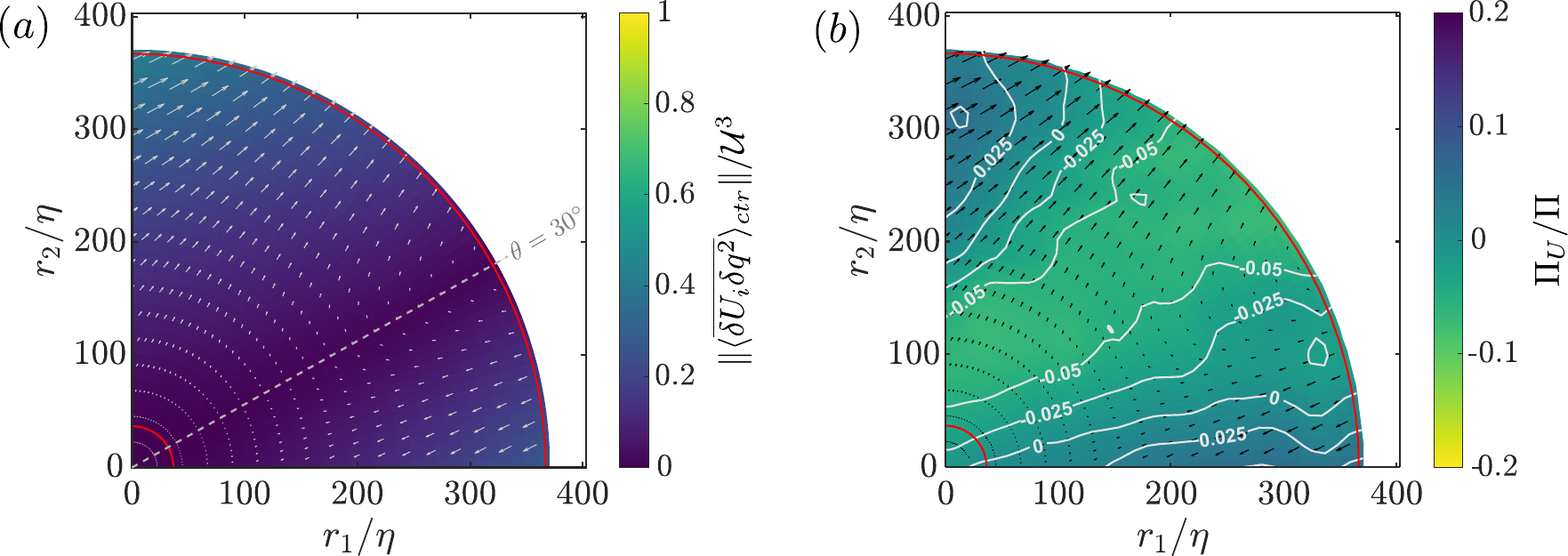}
\caption{\label{fig:khh_linear_flux_7.17Hz} (a) Normalized linear energy flux corresponding to the 7.17~Hz propeller frequency. (b) Ratio between the linear energy transfer and the nonlinear energy transfer of $\overline{\delta q^2}$. The innermost red circle denotes the size of the Taylor microscale, while the outermost corresponds to the integral length scale. Only half of the vectors are plotted.}
\end{figure}
We present in Figure~(\ref{fig:khh_linear_flux_7.17Hz})~(a) the normalized linear flux $\delta U_i \overline{\delta q^2}$ using the same colorbar range as the nonlinear flux for direct comparison (Fig.~\ref{fig:water_khh_third_order_7.17Hz}~(a)). The vector field evidences that the mean flow distributes $\overline{\delta q^2}$ in a different way than the turbulent fluctuations. While the nonlinear flux is predominantly directed from large to small scales across the entire scale space (Fig.~\ref{fig:water_khh_third_order_7.17Hz}~(a)), the linear flux exhibits a negligible magnitude around $\theta \approx 30^\circ$, it is directed from large-to-small scales for $\theta \lessapprox 30^\circ$, and shows the opposite trend for $\theta \gtrapprox 30^\circ$. In addition to these directional differences, the overall magnitude of the linear flux is significantly smaller than that of its nonlinear counterpart. To assess its contribution to the energy balance, we compute its divergence, $\Pi_U = (1/4) (\partial \langle \delta U_k \overline{\delta q^2} \rangle_{ctr} / \partial r_k)$ and compare it with the nonlinear transfer $\Pi$ in Fig.~\ref{fig:khh_linear_flux_7.17Hz}~(b). We find that, even locally, the energy transfer associated with the mean flow is generally below 5\% of that produced by turbulent fluctuations. This is in contrast with the results of \citet{knutsen_inter-scale_2020} for a von Kármán flow, where comparable magnitudes of $\Pi$ and $\Pi_U$ were reported locally. Given that our setup can be regarded as a ``generalized von Kármán'' flow \citep{wang2025lagrangian}, these results suggest that increasing the symmetry of the forcing, by incorporating additional forcing directions, effectively mitigates the impact of mean flow on the interscale energy transfer.

\subsection{Orientation-averaged statistics}
Having discussed the interscale energy distribution and transfer, and highlighted their differences from the classical energy cascade picture, we now consider orientation-averaged statistics to assess whether our flow remains consistent with the classical predictions derived under assumptions of local homogeneity and isotropy. To this end, we employ the spherical average introduced in Sec.~\ref{sec:results_lengthscale}, which allows us to examine the average behavior across structures of the same size, given by the scalar separation $r = \lVert \bm{r} \rVert$.

\begin{figure}
\begin{center}
\includegraphics[width=.55\textwidth]{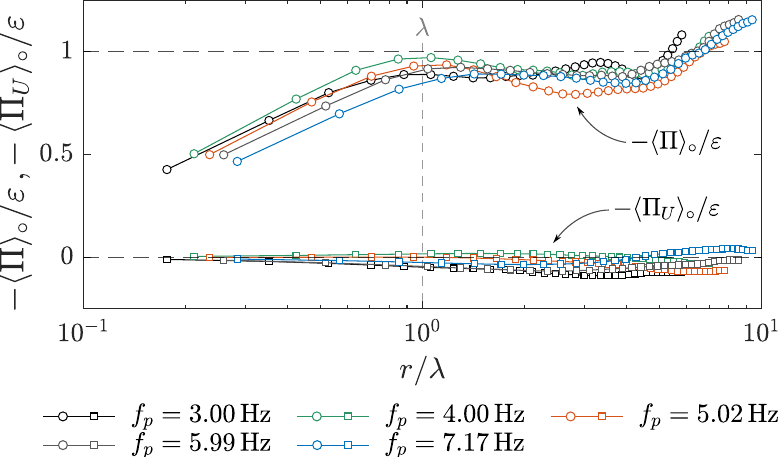}
\end{center}
\caption{\label{fig:Pi_PiU_saa} Spherically averaged linear and nonlinear energy transfers normalized by the dissipation rate, $-\langle \Pi_U \rangle_{\circ}/\varepsilon$ and $-\langle \Pi \rangle_{\circ}/\varepsilon$, for the different propeller frequencies. Circular and square markers correspond to $\Pi$ and $\Pi_U$, respectively.}
\end{figure}

In first place, we present in Fig.~\ref{fig:Pi_PiU_saa} the spherically averaged linear and nonlinear transfer terms, normalized by the dissipation rate obtained in Sec.~\ref{sec:results_dissipation}. As expected from the local analysis in the previous section, the contribution of the linear energy transfer is negligible for all scales and propeller frequencies. In contrast, the nonlinear energy transfer increases in magnitude with separation up to $r \approx \lambda$, where it remains approximately constant at a value of the order of the dissipation rate ($\langle \Pi \rangle_\circ \approx -0.9 \,\varepsilon$) for all propeller frequencies. These results are consistent with the interpretation of the Taylor microscale as marking the limit of the inertial subrange \citep{wang2025energy}. Above this scale, the assumption of local dynamic equilibrium considered by Kolmogorov (Eq.~\ref{eq:kolmogorov_balance}) is approximately satisfied over a significant range of scales. At the largest scales considered, a further increase in the nonlinear energy transfer is observed, which we attribute to the influence of the large-scale forcing.

We turn now to the evaluation of the expressions deduced for second- and third-order structure functions. To this end, we estimate the longitudinal and transverse second-order structure functions by considering, respectively,
\begin{align}
    \langle \overline{\delta q^2_r} \rangle_{ctr,\circ} (r) &= \langle \overline{\delta u_i \delta u_j} \rangle_{ctr,\circ} \hat{r}_i \hat{r}_j \label{eq:second_order_long} \\
    \langle \overline{\delta q^2_{\theta}} \rangle_{ctr,\circ} (r) &= \langle \overline{\delta u_i \delta u_j} \rangle_{ctr,\circ} \hat{\theta}_i \hat{\theta}_j \label{eq:second_order_trans}
\end{align}

\noindent where $\bm{\hat{r}}$ and $\bm{\hat{\theta}} $ are the unit vectors in the radial and polar directions (see Fig.~\ref{fig:setup}~(c)). Similarly, we project the third-order structure function in the longitudinal direction to obtain
\begin{align}
    \langle \overline{\delta u_r \delta q^2} \rangle _{ctr,\circ} (r)= \langle \overline{\delta u_i \delta q^2} \rangle_{ctr,\circ} \, \hat{r}_i \label{eq:third_order}
\end{align}

\begin{figure*}
\begin{center}
\includegraphics[width=\textwidth]{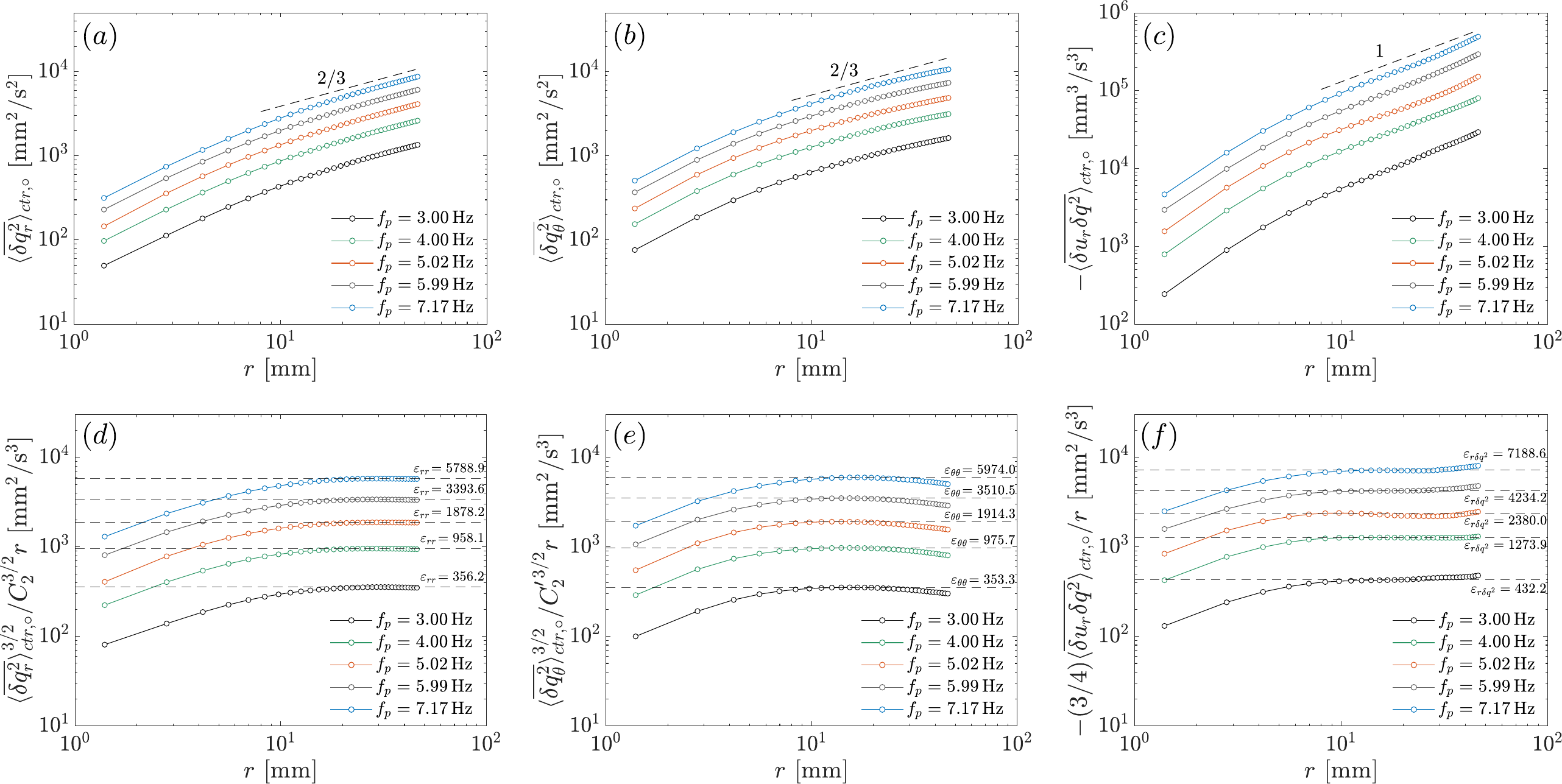}
\end{center}
\caption{\label{fig:water_second-third_saa} Spherically averaged structure functions. Second-order longitudinal (a),  second-order transverse (b), third-order longitudinal (c); and dissipation rate estimates from second-order longitudinal structure functions (d), second-order transverse structure functions (e), and third-order structure functions (f) for the different propeller frequencies.}
\end{figure*}

We recall that, according to \citet{kolmogorov1991local}, a consequence of the dynamic equilibrium is that, in the inertial subrange the statistics are determined by the dissipation rate alone. This implies that within this range the second-order structure functions should satisfy

\begin{align}
    \langle \overline{\delta q^2_r} \rangle_{ctr,\circ} (r) &= C_2 (\varepsilon r)^{2/3} \label{eq:twothirds_long} \\
    \langle \overline{\delta q^2_{\theta}} \rangle_{ctr,\circ} (r) &= C_2^{\, \prime} (\varepsilon r)^{2/3} \label{eq:twothirds_trans}
\end{align}

\noindent where $C_2$ and $C_2^\prime$ are constants expected to be universal \citep{Pope2000a}. Specifically, $C_2 = 2.12$ for $Re_\lambda \gtrapprox 100 $ \citep{sreenivasan_universality_1995} and, if isotropy holds, $C_2 ^{\, \prime} = \frac{4}{3} C_2$ \citep{Pope2000a}. By plotting $\langle \overline{\delta q^2_r} \rangle_{ctr,\circ}$ and $\langle \overline{\delta q^2_{\theta}} \rangle_{ctr,\circ}$ as functions of the separation distance for the different propeller frequencies in Fig.~\ref{fig:water_second-third_saa}~(a) and (b), we observe that they appear compatible with a $2/3$ slope for an intermediate range of scales across all forcing conditions.

Considering now the third-order structure functions, as previously mentioned, a consequence of local dynamic equilibrium and local isotropy is that, in the inertial subrange,
\begin{align}
    \langle \overline{\delta u_r \delta q^2} \rangle _{ctr,\circ} = - \frac{4}{3} \varepsilon r \label{eq:fourfifhts}
\end{align}
\noindent which is equivalent to the four-fifths law, where the proportionality constant in this case can be determined analytically under conditions of homogeneity and isotropy \citep{antonia1997analogy,gomes2015energy}. The spherically averaged third-order structure functions plotted in Fig.~\ref{fig:water_second-third_saa}~(c) suggest a behavior consistent with the predicted linear scaling, $\sim r^1$.
To verify the validity of the classical scalings and simultaneously estimate the dissipation rate, we compute:
\begin{align}
    \varepsilon_{rr} &= \max_{r \in r_{IR}} \frac{\langle \overline{ \delta q_r^2} \rangle_{ctr,\circ} ^{3/2}}{C_2 ^{3/2} r} \label{eq:epsilon_rr} \\
    \varepsilon_{\theta \theta} &= \max_{r \in r_{IR}} \frac{\langle \overline{\delta q_{\theta}^2} \rangle_{ctr,\circ} ^{3/2}}{C_2 ^{\, \prime \, 3/2}r} \label{eq:epsilon_tt} \\
    \varepsilon_{rq^2} &= \max_{r \in r_{IR}} -\frac{3}{4} \frac{1}{r} \langle \overline{\delta u_r \delta q^2} \rangle_{ctr,\circ} \label{eq:epsilon_rq2}
\end{align}

\noindent where $r_{IR} = [10, \, 22] \, \mathrm{mm}$ denotes the range of separations within the inertial subrange.
\begin{figure}
\begin{center}
\includegraphics[width=.45\textwidth]{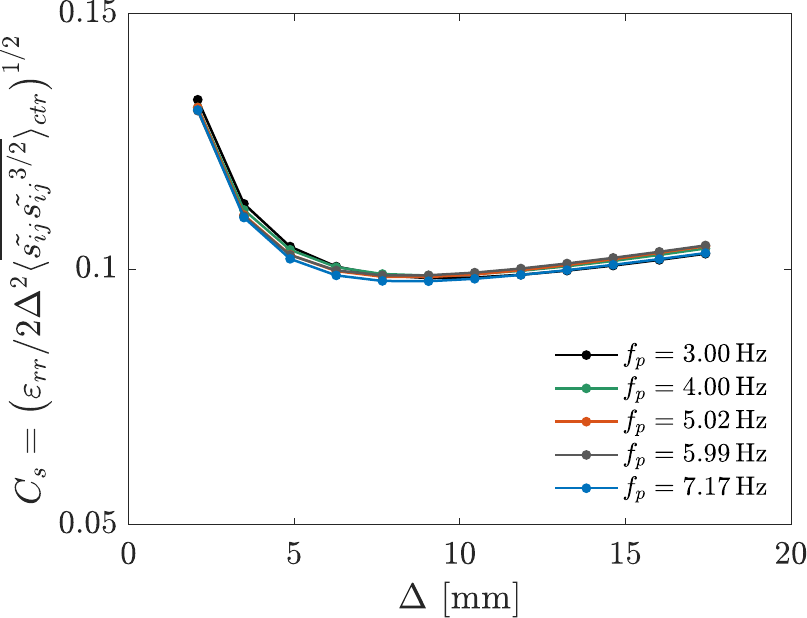}
\end{center}
\caption{\label{fig:lepiv-filtersize} Verification of the Smagorinsky constant $C_s$ for different filter sizes $\Delta$ and all experimental conditions.}
\end{figure}
Figures~\ref{fig:water_second-third_saa}~(d)-(f) illustrate the determination of dissipation rates from the longitudinal ($\varepsilon_{rr}$), transverse ($\varepsilon_{\theta \theta}$), and third-order ($\varepsilon_{rq^2}$) compensated structure functions, with the resulting values presented in Fig.~\ref{fig:dissipation_rates}~(a). The observed plateaus denote the separations for which the second- and third-order structure functions agree with Eqs.~(\ref{eq:twothirds_long}), (\ref{eq:twothirds_trans}) and (\ref{eq:fourfifhts}),  scaling as $\sim r^{2/3}$ and $\sim r^{1}$, as suggested by Fig.~\ref{fig:water_second-third_saa}~(a)-(c). We note that the plateaus extend further into the large-scale separations for the longitudinal second-order structure functions (Fig.~\ref{fig:water_second-third_saa}~(d)) than for the transverse (Fig.~\ref{fig:water_second-third_saa}~(e)), where the compensated values slightly decay at the largest measured scales. This indicates a higher sensitivity of transverse structure functions to the external forcing. However, we consider it a minor deviation since it does not significantly influence the energy distribution observed in Fig.~\ref{fig:khh_second_order_7.17Hz}, nor the isotropy indices discussed in Sec.~\ref{subsec:flow_quality} (Fig.~\ref{fig:flow-quality}~(b)).

From the values of the plateaus, we estimate the dissipation rates for the different structure functions across all experimental conditions, as shown by the dashed lines in Figs.~\ref{fig:water_second-third_saa}~(d)-(f). We observe that the values obtained for the second-order longitudinal and transverse structure functions agree within 4\%, as expected from the high degree of isotropy of the energy distribution observed in our flow (Fig.~\ref{fig:khh_second_order_7.17Hz}). Conversely, dissipation rates estimated by third-order structure functions are approximately 20-35\% larger than those obtained from second-order structure functions. We attribute these differences to the observed anisotropy of $\langle \overline{ \delta u_i \delta q^2 }\rangle_{ctr}$, which violates the isotropy assumptions leading to the $-4/3$ constant. Note that the two-thirds law is based only on the local dynamic equilibrium assumption (Eq.~\ref{eq:kolmogorov_balance}), resulting from local homogeneity and high Reynolds numbers. This equilibrium is indeed satisfied in our flow, as shown in Fig.~\ref{fig:Pi_PiU_saa}, thereby justifying the observed $r^{2/3}$ scaling. Furthermore, the isotropy of $\langle \overline{\delta q^2} \rangle_{ctr}$ observed in the previous section supports the selected values of $C_2^\prime$. On the other hand, as noted in Refs.~\citep{vassilicos2015dissipation,gomes2015energy}, the scaling expression for the third-order structure function (Eq.~\ref{eq:third_order_durdq2}) is obtained by additionally assuming isotropy of the third-order structure function. Because $\langle \overline{\delta u_i \delta q^2} \rangle_{ctr}$ is not perfectly isotropic in our flow, deviations from the isotropic $-4/3$ constant are to be expected, leading to discrepancies in the determination of $\varepsilon$. This observation is consistent with the results of \citet{gomes2015energy}, who reported a $-16/9$ factor in their anisotropic third-order structure functions, rather than the $-4/3$ constant expected from Richardson-Kolmogorov phenomenology.

As an additional verification of the dissipation rate, we consider our prior estimate based on the large-eddy PIV method (Sec.~\ref{sec:results_dissipation}), and evaluate the selection of both the filter size $\Delta$ and the Smagorinsky constant $C_s$. To this end, we follow the approach of \citet{lawson_unsteady_2022} and determine the value of $C_s$ that matches, in the HIT region, the dissipation estimated from the longitudinal second-order structure function $\varepsilon_{rr}$. Employing Eq.~(\ref{eq:lepiv}), we express the Smagorinsky constant as $C_s = (\varepsilon_{rr} / 2\Delta ^2 \langle\overline{(\tilde{s}_{ij}\tilde{s}_{ij}) ^{3/2} }\rangle _{ctr})^{1/2}$. Then, for each propeller frequency, $\varepsilon_{rr}$ is taken from the structure function estimation while $\langle\overline{(\tilde{s}_{ij}\tilde{s}_{ij}) ^{3/2} }\rangle _{ctr}$ is computed by filtering the velocity fields across varying filter widths $\Delta$.

Figure~\ref{fig:lepiv-filtersize} presents the values of $C_s$ for different filter sizes $\Delta$ and all experimental conditions. We observe that our selection $\Delta = 10.45$ mm lies within a range where $C_s$ is nearly independent of the filter size. This shows that the energy flux is scale-invariant for this $\Delta$, a behavior expected for separations within the inertial subrange. Additionally, the range of scales where $C_s$ remains nearly constant is compatible with the range where $-\Pi\approx \varepsilon$ as well as with the interval of validity of the $r^{2/3}$ and $r^{1}$ scalings (Fig.~\ref{fig:water_second-third_saa}~(d)-(f)), thereby confirming the inertial subrange. Since the value of $C_s$ is approximately equal to 0.1 in this plateau region, we note that the dissipation rates obtained from the two-thirds law and the large-eddy PIV method match closely across all tested forcing conditions if standard values of $C_s=0.1$ and $C_2 = 2.12$ are adopted.

\section{Temporal perspective: Eulerian frequency spectra} \label{sec:eulerian_spectra}
\begin{figure}
\includegraphics[width=.85\textwidth]{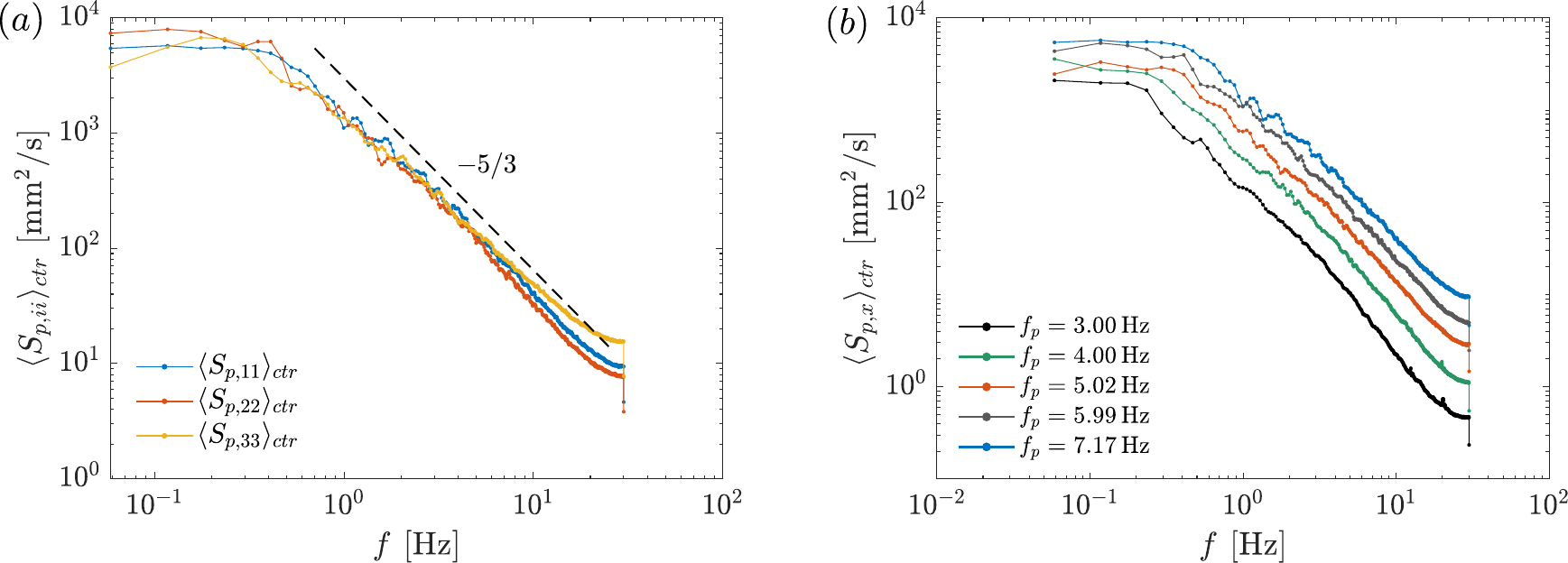}
\caption{\label{fig:eulerian_spectra} (a) Frequency spectra averaged over the central region corresponding to the $f_p=7.17$ Hz forcing condition and the different velocity components. (b) Frequency spectra corresponding to the horizontal velocity component and all forcing conditions.}
\end{figure}

The previous analysis, based on structure functions, involves the time averaging of spatial statistics. This hides the temporal variability, which should also reflect the imprint of the turbulent dynamics. To address this aspect, in this section we analyze the Eulerian frequency spectra obtained from temporal velocity signals measured by stationary probes.

Typically, measurements performed in time in fundamental studies of turbulence are directly interpreted in the spatial domain by means of the Taylor's frozen turbulence hypothesis \citep{taylor1938spectrum}. This hypothesis assumes that, if the mean velocity $U$ is much larger than the turbulent velocity $u^{\prime}$, the temporal response of the flow recorded at a point by a fixed probe is the consequence of the passage of an unchanging pattern of turbulent motion convected uniformly at velocity $U$. This concept has been of great utility for exploring the interscale energy dynamics and its consequences from temporal measurements at fixed points in grid turbulence \citep[e.g.][]{mora2019energy}, and requires $u^{\prime}/U \ll 1$. Unlike wind tunnels, HIT chambers present a weak mean flow as discussed in Sec.~\ref{subsec:flow_quality}, thereby invalidating the use of Taylor's frozen hypothesis. Under these circumstances, where the mean flow is weak and statistical isotropy holds, \citet{tennekes1975eulerian} proposed that the signal recorded by a fixed probe is dominated by the advection of small-scale eddies by the large scales of the flow. This phenomenology, also called the random sweeping hypothesis \citep{sagaut2018homogeneous}, leads to the following form for the Eulerian frequency spectrum for an intermediate range of scales,
\begin{equation}
E(\omega) = B  \, (\sqrt{3} \, \mathcal{U})^{2/3} \varepsilon^{2/3} \omega^{-5/3} \label{eq:eulerian_spectrum}
\end{equation}

\noindent where $\omega = 2\pi f$ is the angular frequency, $\mathcal{U}$ is the velocity scale of the energy-containing eddies, $B$ is a constant, and the factor $\sqrt{3}$ arises for compatibility with the different velocity scale adopted by \citet{tennekes1975eulerian}. It should be noted that this range is not determined uniquely by dissipation as stated by \citet{kolmogorov1991local}, but by both the integral velocity scale and the dissipation rate, and it is thus termed the inertial-convective range by \citet{tennekes1975eulerian}.

\begin{figure*}
\includegraphics[width=0.95\textwidth]{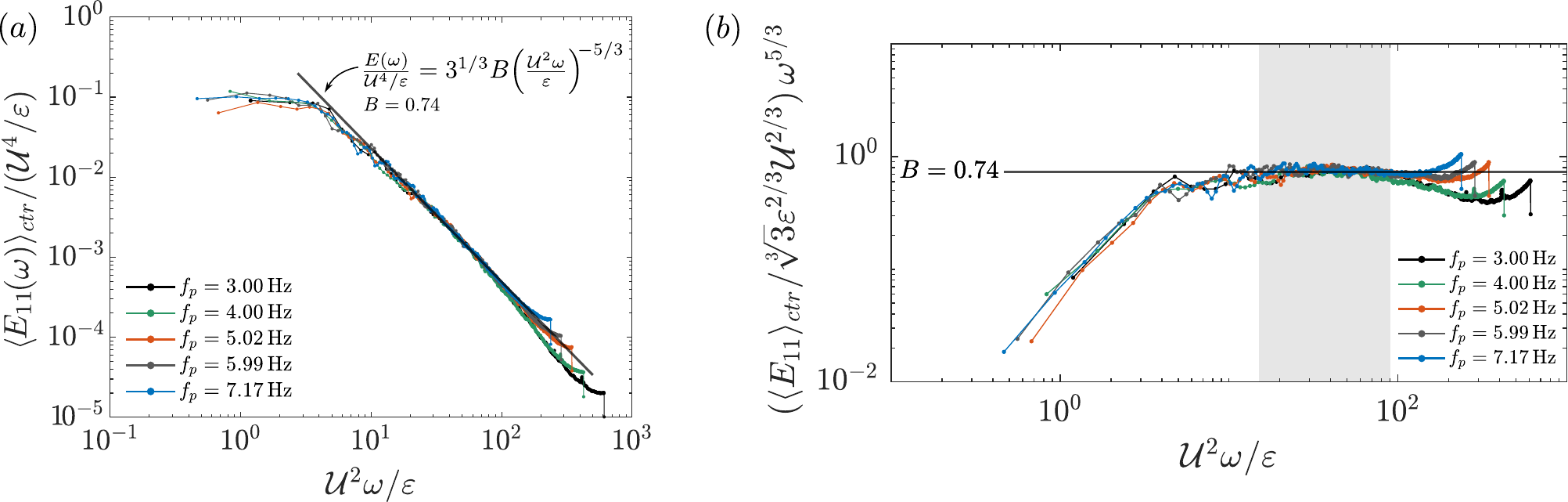}
\caption{\label{fig:eulerian_spectra_normalized} (a) Eulerian frequency spectra corresponding to the horizontal velocity component and all forcing conditions, normalized according to Eq.~(\ref{eq:eulerian_spectrum_normalized}). The gray line shows the \citet{tennekes1975eulerian} deduction for the inertial-convective range with the value of the constant $B$ obtained from our experimental data. (b) Compensated Eulerian spectra corresponding to the horizontal velocity component and all forcing conditions. The shaded region denotes the range of frequencies considered for fitting $B$.}
\end{figure*}

We evaluate the applicability of the random sweeping hypothesis to describe the turbulence generated by the spherical forcing by analyzing measurements performed with the same velocimetry setup described in Sec.~\ref{sec:methods}, but at a sampling frequency of 60 Hz. We compute the frequency spectra for each velocity component at every spatial point and analyze the results spatially averaged over the central region ($ |x|/D \leq 0.3$ and $|y|/D \leq 0.3$). Figure~\ref{fig:eulerian_spectra}~(a) presents the spectra $\langle S_{p,ii} (f)\rangle_{ctr}$ for the $f_p = 7.17$ Hz forcing case, where we observe a considerable power-law range compatible with a -5/3 slope. As a consequence of the isotropy of the flow, the curves of the different velocity components collapse over the full range of scales, except for the high-frequency end of the out-of-plane component, which appears to be more affected by noise. Figure~\ref{fig:eulerian_spectra}~(b) shows the spectra corresponding to the horizontal component across all experimental cases, illustrating the systematic increase in energy resulting from higher forcing speeds.

To assess the applicability of the Tennekes spectrum to our experimental data, we analyze the normalized form of Eq.~(\ref{eq:eulerian_spectrum}),
\begin{equation}
    \frac{E(\omega)}{\mathcal{U}^4/\varepsilon} = \sqrt[3]{3} B \Bigg( \frac{\mathcal{U}^2\omega}{\varepsilon} \Bigg)^{-5/3} \label{eq:eulerian_spectrum_normalized}
\end{equation}

We then scale our results to obtain the angular frequency spectra $\langle E_{ii}(\omega)\rangle_{ctr}$ and normalize them using as dissipation and velocity scales the values obtained from the 1~Hz measurements in Secs.~\ref{sec:results_dissipation} and \ref{sec:results_lengthscale}, respectively. The results, presented in Figure~\ref{fig:eulerian_spectra_normalized}~(a), indicate that the spectra corresponding to different forcing conditions collapse onto a single curve. To examine the spectral slope more clearly, we also present the compensated spectra in Fig.~\ref{fig:eulerian_spectra_normalized}~(b), where our results are consistent with a $-5/3$ power law over approximately a decade. A fit of the constant $B$ over this range, indicated in gray, yields $B=0.74$.

Originally, \citet{tennekes1975eulerian} suggested that $B$ should be of order one, while subsequent kinematic simulations of \citet{fung1992kinematic} obtained a value of 0.54. Since then, experimental determinations of $B$ have been reported for various turbulence generation mechanisms, mainly employing oscillating grids and randomly actuated jet arrays. For oscillating grid turbulence, \citet{desilva1994oscillating} originally reported a relatively large value of $B=5.54$. However, subsequent refinements based on independent estimates of the dissipation rate led to significantly lower values, within the range $0.49 \leq B \leq 0.52$ \citep{kit1995experimental}. In contrast, experiments using randomly actuated synthetic jets yield systematically lower estimates, with $B=0.23$ \citep{Variano2008} and $B=0.12$-$0.15$ \citep{johnson2018turbulent}. Among the available studies, perhaps the experimental evaluation of $B$ under the most isotropic conditions is given by \citet{cazaubiel2021three} for a flow generated by magnetic particles, obtaining $B=0.64 \pm 0.15$. Our measured value of $B=0.74$ is in good overall agreement with previously reported results, particularly with those of \citet{cazaubiel2021three}, obtained under nearly isotropic conditions, with remaining differences likely reflecting variations in the degree of isotropy introduced by the respective forcing mechanisms.

\section{Global perspective: scaling of main turbulent statistics}
\label{sec:scaling}
In the preceding sections, we analyzed a series of statistics characterizing the interscale energy transfer and observed their connection with the main flow scales, such as $\mathcal{\varepsilon}$, $\mathcal{U}$ and $\lambda$, within the frameworks of the Richardson-Kolmogorov phenomenology and the random sweeping hypothesis for spatial and temporal statistics, respectively. Specifically, the spatial analysis revealed an inertial subrange dominated by $\varepsilon$ and bounded by $\lambda$, whereas the temporal analysis showed an inertial-convective range dominated by $\mathcal{U}$ and $\varepsilon$. In this section we investigate whether these flow scales can be linked to external experimental variables. A thorough understanding of the role of external forcing is essential for designing strategies that produce favorable flow conditions across different applications. 

We begin by considering an energy balance in the entire volume $\mathcal{V}$ of a generic HIT chamber, with $n_p$ propellers operating at a fixed rotational frequency $f_p$, and an external heat sink maintaining the fluid temperature constant. Under statistically stationary conditions, the energy balance is expressed as:
\begin{equation} 
n_p P = \int _{\mathcal{V}} \rho \, \varepsilon _{tot} \, \mathrm{d}\mathcal{V}
\label{eq:balance1}
\end{equation}
\noindent where $P$ is the input power of each propeller, $\rho$ is the fluid density, and the dissipation of kinetic energy is $\varepsilon _{tot} \approx \varepsilon = 2 \nu \overline{ s_{ij} s_{ij} }$, where $s_{ij} = 1/2 (\partial u_i / \partial x_j + \partial u_j / \partial x_i)$ is the fluctuating strain rate tensor. For the sake of simplicity we analyze an average rate of dissipation over the chamber volume, $\langle \varepsilon \rangle _{\mathcal{V}}  = (1/\mathcal{V}) \int _{\mathcal{V}} \varepsilon_{tot} d\mathcal{V} $, which from Eq.~(\ref{eq:balance1}) results in,
\begin{equation}
 \langle \varepsilon \rangle _{\mathcal{V}} = \frac{n_p P}{\rho \mathcal{V}} \label{eq:balance3}
\end{equation}
We consider the dependence of the power input by each propeller on the external experimental variables. A dimensional analysis of $P$ shows that the non-dimensional input power is a function of the Reynolds number of the propeller $Re_I = f_p D^2/ \nu $ and the geometrical parameters characterizing the chamber and the propellers,
\begin{equation}
\frac{P}{\rho f_p^3 D^5} =  g_1(Re_I  , \text{geom. parameters}) \label{eq:P_dimanalysis}
\end{equation}
where $g_1(\cdot)$ is a scaling function \citep{mordant1997characterization,paul2004handbook}. The shape of $g_1(Re_I)$ can be studied by measuring the electric consumption or the torque of the motor. This is conventionally done in the analysis of von Kármán swirling flows \citep{mordant1997characterization,ravelet_supercritical_2008,burnishev_torque_2014} and stirred tanks \citep{sirasitthichoke2017power,steiros2017effect}. 
For high enough Reynolds numbers, the dimensionless input power becomes independent of viscosity, i.e. $g_1(Re_I) \sim \text{constant}$ ($=K_{P}$). The average dissipation rate can then be expressed as,
\begin{equation}
\langle \varepsilon \rangle _{\mathcal{V}} = K_{P} \frac{n_p D^5}{\mathcal{V}} f_p^3  \label{eq:dissipation_turbulent}
\end{equation}

Although Eq.~(\ref{eq:dissipation_turbulent}) shows the dependence of $\varepsilon$ averaged over the whole chamber, we consider as a surrogate the estimate of dissipation computed in Sec.~\ref{sec:results_dissipation}, which only describes the flow behavior in the chamber center. In this central region, we observed that this local estimate correctly captures the average energy transfer rate within the inertial subrange of second-order structure functions (Sec.~\ref{sec:khh}) and properly scales the Eulerian spectra along with $\mathcal{U}$ (Sec.~\ref{sec:eulerian_spectra}). We first note that the cubic dependence obtained analytically in Eq.~(\ref{eq:dissipation_turbulent}) is in good agreement with the best-fit curve shown in Fig.~\ref{fig:dissipation_rates}~(a). To further assess the robustness of this expression, we collect data from a series of studies characterizing propeller-driven HIT chambers with continuous forcing. The available literature covers a wide range of experimental conditions, including diverse propeller and chamber geometries, varying number and type of propellers, and working fluids. Since these studies typically report the dependence of turbulent scales on the propeller frequency, we focus on evaluating the ability of the analytical expression to capture the scaling of these variables with propeller speed. To facilitate the comparison between devices employing fluids of different viscosities, we rewrite Eq.~(\ref{eq:dissipation_turbulent}) to express the propeller speed dependence  in terms of $Re_I = f_p D / \nu$. Additionally, we analyze the normalized dissipation $\tilde \varepsilon$ as a function of $Re_I$ to highlight the expected power-law dependence,
\begin{align}
    \tilde\varepsilon = \frac{\langle \varepsilon \rangle _{ctr}}{K_P^{ctr} n_p  \nu^3/\mathcal{V}D} = Re_I ^3
\end{align}

\begin{figure*}
\includegraphics[width=0.95\textwidth]{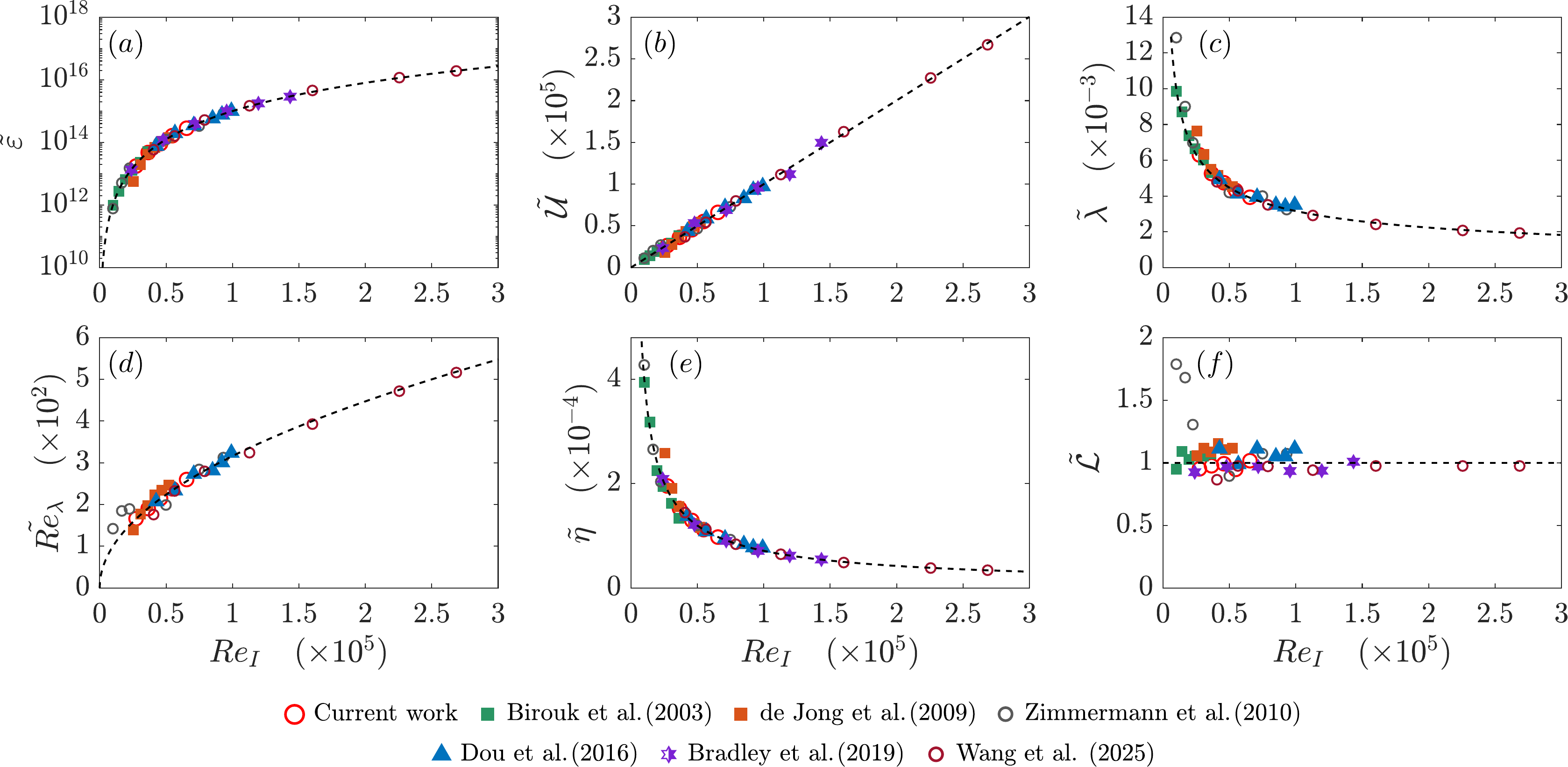}
\caption{\label{fig:scaling_all} Scaling of normalized variables with $Re _I$: dissipation rate (a), velocity scale (b), Taylor microscale (c), Taylor Reynolds number (d), Kolmogorov length scale (e), and integral length scale (f). Symbols represent the chamber geometry or number of propellers: icosahedron ($n_p=12$, circles), cube ($n_p=8$, squares), truncated icosahedron ($n_p=20$, triangles) and sphere ($n_p=4$, stars). Symbol fill represent the test fluid: water (hollow), air (filled) and helium (semi-filled). The Taylor microscale from \citet{bradley_measurement_2019} and Taylor Reynolds number data from \citet{bradley_measurement_2019} and \citet{birouk_attempt_2003} were excluded from the analysis due to inconsistencies in their original evaluation.}
\end{figure*}

We fit the experimental constant for each study and observe that the obtained values are within the same order of magnitude, $1.8 \times 10^{-2} \leq K_P^{ctr} \leq 8.4 \times 10^{-2}$ (see Table~\ref{tab:table2}). The only exception is the work of \citet{birouk_attempt_2003}, where the authors assumed dynamic equilibrium $\varepsilon = k^{3/2}/\mathcal{L}$. In that case, the relatively small integral length scale reported ($\mathcal{L} \approx 9 \, \mathrm{mm}$) and the larger velocity scale, both in comparison with the similar configuration of \citet{DeJong2009}, likely lead to an overestimation of $K_P^{ctr}$. Despite this difference, $\tilde\varepsilon$ shows an excellent agreement with the $Re_I^3$ scaling across all devices considered, in line with the analytical prediction (Fig.~\ref{fig:scaling_all}~(a)). It should be noted that all the compiled data correspond to dissipation rates representative of the chamber center, as in the present study. Moreover, the robustness of the analytical prediction is supported by the fact that $\varepsilon$ is estimated using different methods across these works, including global scaling \citep{birouk_attempt_2003}, structure functions \citep{Zimmermann2010,Dou2016a,wang2025lagrangian}, and direct computation from velocity gradients \citep{bradley_measurement_2019,verwey_dissipation_2022}.

\begin{table}[b]
\caption{\label{tab:table2}
Main characteristics and fitted scaling constants of the propeller-driven HIT chambers used for comparison: number of propellers $n_p$, chamber geometry, volume $\mathcal{V}$, energy injection constant $K_P^{ctr}$, integral velocity scale constant $K_\mathcal{U}$, and dissipation constant $C_\varepsilon$.}
\begin{ruledtabular}
\begin{tabular}{lccccccc}
      Work & Fluid & Chamber geometry &  $n_p$ & $ \mathcal{V} $ [m\textsuperscript{3}] & $K_P^{ctr} \, (\times 10^{-2})$ & $K_\mathcal{U} \, (\times 10^{-2})$ & $C_\varepsilon$ \\
     \hline
     \citet{birouk_attempt_2003} & Air & Cube & 8 & 0.064 & $97.8 \pm 24.3$ & $6.7 \pm 0.6$ & $1.84$ \\
     \citet{DeJong2009} & Air & Cube & 8 & 0.064 & $6.9 \pm 1.4$ & $5.1 \pm 1.0$ & $1.84$ \\
     \citet{Zimmermann2010} & Water & Icosahedron & 12 & 0.140 & $4.9 \pm 1.2$ & $3.8 \pm 0.5$ & $1.00$ \\
     \citet{Dou2016a} & Air & Truncated Icosahedron & 20 & 0.483 & $5.3 \pm 0.4$ & $3.8 \pm 0.2$ & $1.84$\\
     \citet{bradley_measurement_2019} & Air & Sphere & 4 & 0.029 & $8.4 \pm 0.8$ & $26.0 \pm 3.0$ & $0.04$\\
     \citet{wang2025lagrangian} & Water & Icosahedron & 12 & 0.140 & $3.7 \pm 0.2$ & $4.9 \pm 0.1$ & $0.71$\\
     Present work & Water & Icosahedron & 12 & 0.140 & $1.8 \pm 0.1$ & $3.4 \pm 0.1$ & $0.43$\\
     \end{tabular}
\end{ruledtabular}
\end{table}

We also investigate the dependence of the integral velocity scale on the experimental parameters. We consider that, in addition to the energy input, a variable quantifying the capacity of the propellers to generate turbulence at the chamber center is required. For this purpose, propellers with geometric designs that convert the input power into a purely axial flow would be more efficient. Based on this reasoning, we follow \citet{Dou2016a} and assume that each propeller generates a jet characterized by a velocity scale $U_p$, and that the TKE at the center of the chamber results from equal contributions of all propellers, $k \sim n_p U_p ^2$. To relate $U_p$ to experimental parameters, we assume the jet crosses an area $A \propto \pi D^2/4$, yielding a volumetric discharge $Q = U_p A$. Analogously to the discussion regarding the input power, dimensional analysis of $Q$ leads to
\begin{equation}
    \frac{Q}{f_p D^3} =  g_2(Re_I, \text{geom. parameters}) \label{eq:Q_dimanalysis}
\end{equation}
where the dimensionless discharge $Q / (f_p D^3)$ is known as the pumping number in the chemical engineering literature \citep{paul2004handbook}. Its dependence on the Reynolds number has been extensively studied in mixing vessel setups for several geometric arrangements, where it is found to be constant at high $Re_I$, i.e. $g_2(Re_I) \approx \text{constant}$ \citep{paul2004handbook}. Consequently, under fully developed turbulent conditions, $U_p = Q/A \propto f_p D$, from which the TKE scales as $k \sim n_p U_p ^2 \propto n_p f_p^2 D^2$. The integral velocity scale then yields
\begin{equation}
\mathcal{U} = \sqrt{2/3 k} = K_\mathcal{U} n_p^{0.5} f_p D
\label{eq:velocity_scale}
\end{equation}
\noindent where $K_\mathcal{U}$ is a constant at high Reynolds numbers. We note that the linear scaling predicted by Eq.~(\ref{eq:velocity_scale}) is in good agreement with our results in Fig.~\ref{fig:length-velocity-scales}~(a), both for the entire FOV and central region. To further assess its validity across different experimental conditions, we fit $K_\mathcal{U}$ for the HIT chambers characterized in the literature and evaluate the normalized velocity scale as a function of the Reynolds number,
\begin{align}
\tilde{\mathcal{U}} = \frac{\mathcal{U}}{K_\mathcal{U} \nu n_p^{0.5}/D} = Re_I
\end{align}
Fitting the constants for the different experimental setups yields values within the range $3.4 \times 10^{-2} \leq K_\mathcal{U} \leq 6.7 \times 10^{-2}$, with the only exception of \citet{bradley_measurement_2019}, who reported unusually large velocities. Moreover, the velocity scale shows excellent agreement with the linear scaling across all considered devices (Fig.~\ref{fig:scaling_all}~(b)).

Given the analytical expressions derived for $\varepsilon$ and $\mathcal{U}$ in Eqs.~(\ref{eq:dissipation_turbulent}) and (\ref{eq:velocity_scale}), the scaling relations with respect to the propeller frequency for the Taylor microscale $\lambda = \sqrt{15 \nu \mathcal{U}^2 / \varepsilon}$, the Taylor Reynolds number $Re_\lambda = \mathcal{U}\lambda/\nu$, and the Kolmogorov length scale $\eta = (\nu^3/\varepsilon)^{1/4}$, are obtained as,
\begin{align}
    \lambda &= \frac{K_\mathcal{U}}{K_p^{ctr\,1/2}} \left( \frac{15 \nu \mathcal{V}}{D^3} \right)^{1/2} f_p^{-1/2} \\
    Re_\lambda &= \frac{K_\mathcal{U}^2}{K_P^{1/2}} \left( \frac{15 n_p \mathcal{V}}{\nu D} \right)^{1/2} f_p ^{1/2}\\
    \eta &= K_P^{ctr\,-1/4} \left( \frac{\nu^3 \mathcal{V}}{n_p D^5} \right)^{1/4} f_p^{-3/4}
\end{align}

Since the experimental data show good agreement with the scaling expressions for the dissipation rate and velocity scale, they are also consistent with the corresponding predictions for $\lambda$, $Re_\lambda$ and $\eta$. To demonstrate this, we plot in Fig.~\ref{fig:scaling_all}~(c)-(e) the normalized variables $\tilde{\lambda} = \lambda / [(K_\mathcal{U}/K_p^{ctr\,1/2})(15 \nu \mathcal{V}/D)^{1/2}] = Re_I^{-1/2}$, $\tilde{Re_\lambda} = Re_\lambda / [(K_\mathcal{U}^2/K_P ^{ctr \, 1/2}) (15 n_p \mathcal{V}/D^3)^{1/2}] = Re_I^{1/2}$ and $\tilde{\eta} = \eta / [K_P^{ctr \, -1/4} (\mathcal{V} D/n_p )^{1/4}] = Re_I^{-3/4}$. 

Finally, we consider the dependence of the integral length scale $\mathcal{L}$ on the forcing conditions. As noted by \citet{mordant1997characterization}, $\mathcal{L}$ is often identified as a characteristic geometrical dimension of the experiment. In this sense, it acts as a fixed parameter that is independent of the forcing, even though its physical meaning is to capture the size of the energy-containing eddies. Accordingly, the authors propose to equate the rate of energy injection, given in our case by $n_p P / \rho \mathcal{V}$, with the rate of energy transfer by structures of integral size, $\mathcal{U}^3 / \mathcal{L}$. Since, under statistically stationary conditions, the rate of energy injection balances the rate of dissipation (Eq.~\ref{eq:balance1}), this leads to a dynamic equilibrium between integral and dissipative scales,
\begin{align}
    \varepsilon = C_\varepsilon \mathcal{U}^3/\mathcal{L} \label{eq:global_scaling}
\end{align}
where the dissipation constant $C_\varepsilon$ is independent of $Re_\lambda$ at sufficiently high Reynolds numbers. Consequently, under fully developed turbulent conditions, substituting our estimates of $\varepsilon$ and $\mathcal{U}$, Eqs.~(\ref{eq:velocity_scale}) and (\ref{eq:dissipation_turbulent}), respectively, yields an integral length scale independent of propeller frequency,
\begin{equation}
\mathcal{L} = C_\varepsilon \frac{K_\mathcal{U}^3}{K_P ^{ctr}} \frac{ \mathcal{V}n_p^{1/2}}{D^2} \, f_p^0 \label{eq:lenghscale_turbulent}
\end{equation}

Although a non-equilibrium dissipation law has been observed in various flow configurations (where $C_\varepsilon$ varies with the Reynolds number even when finite Reynolds number effects are negligible), stationary turbulent flows are expected to follow the conventional equilibrium scaling \cite{vassilicos2015dissipation}. Consequently, for propeller-driven HIT chambers at high-enough $Re_\lambda$, we expect $C_\varepsilon$ to be constant, however with a value depending on the forcing conditions. To evaluate if the geometrical variations between HIT chambers can introduce differences in the value of $C_\varepsilon$, independent determinations of the dissipation rate, integral length scale and velocity scale are required. However, most existing studies on propeller-driven HIT chambers assume dynamic equilibrium and estimate either the dissipation rate or the integral length scale, by setting $\varepsilon = k^{3/2}/\mathcal{L}$ (which implies $C_\varepsilon = (3/2)^{3/2} \approx 1.84$) \citep{birouk_attempt_2003,DeJong2009,Dou2016a}, or $\varepsilon = u^{\prime \, 3}/\mathcal{L}$ ($C_\varepsilon = 1$) \citep{Zimmermann2010}. To the best of our knowledge, only the present work and those of \citet{bradley_measurement_2019} and \citet{wang2025lagrangian} provide independent estimates of all three quantities, thereby permitting a direct evaluation of $C_\varepsilon$. The data presented by \citet{bradley_measurement_2019} lead to approximately constant, but very small values of $C_\varepsilon \approx 0.04$ (Table~\ref{tab:table2}), resulting from their unusually large velocity scales, as previously noted. Conversely, our data yield approximately constant values of $C_\varepsilon \approx 0.43$ and $0.53$, when evaluated using dissipation rates derived from second- and third-order structure functions, respectively. These values are in remarkable agreement with the asymptotic value $C_\varepsilon \approx 0.5$ observed at high Reynolds numbers in direct numerical simulations (DNS) of forced isotropic turbulence \citep[see e.g.,][]{mccomb2015energy}. On the other hand, \citet{wang2025lagrangian} obtained $C_\varepsilon \approx 0.71$, independent of the Reynolds number. The discrepancy between their result and ours may reflect the sensitivity of $C_\varepsilon$ to the forcing mechanism, as their device shares the same chamber geometry but differs in propeller design and diameter. It is also worth noting that, although \citet{verwey_dissipation_2022} specifically aim to estimate $C_\varepsilon$ in a propeller-driven HIT chamber, the Reynolds numbers achieved are not sufficiently high to reach the asymptotic regime where $C_\varepsilon$ becomes independent of Reynolds number, and therefore their results cannot be directly compared with the present analysis.

Despite aforementioned considerations, we evaluate the independence of the integral length scale from the propeller frequency across the different studies by plotting the normalized length scale $\tilde{\mathcal{L}} = \mathcal{L} / [(C_\varepsilon K_\mathcal{U}^3 / K_P^{ctr}) n_p^{1/2} (\mathcal{V}/D^2)]$ as a function of $Re_I$. To this end, we employ the (assumed or estimated) values of $C_\varepsilon$ for each study (Table~\ref{tab:table2}) along with the values of $K_P^{ctr}$ and $K_\mathcal{U}$ obtained previously. As expected from the cubic and linear scaling behavior of the dissipation and velocity scales already verified, the integral length scale is shown to be independent of the propeller frequency across the various experimental configurations. Only the work of \citet{Zimmermann2010} exhibits a significant scatter, where the values were computed employing Eq.~(\ref{eq:global_scaling}) with $C_\varepsilon =1$. Although they report these values, the authors note that evaluating the integral length scale by integrating velocity autocorrelation functions yields a constant value of $\approx 45$ mm, independent of Reynolds numbers, and attribute the observed discrepancies to the Reynolds number dependence of $C_\varepsilon$.

\section{Conclusions} \label{sec:conclusions}
In this work, we investigate previously unexplored aspects of a propeller-driven HIT chamber related to interscale energy dynamics, through a combined analysis of spatial, temporal, and global statistics. We first examined spatial statistics, leveraging the Kármán-Howarth-Monin analysis to study the isotropy of second- and third-order structure functions and the behavior of the energy cascade in scale space. We observe that, in contrast to other forcing strategies, spherical forcing produces a forward energy cascade throughout scale space. While this may appear trivial, prior studies in other canonical flow configurations have shown the coexistence of inverse and forward energy cascades in scale space. We find that for our setup the energy distribution across scales is highly isotropic, and its spherical average agrees with the Kolmogorov's two-thirds law. The energy flux is predominantly radial and directed from large to small scales, in agreement with the Richardson-Kolmogorov energy cascade picture. However, its magnitude exhibits a directional dependence, denoting a higher sensitivity to residual forcing asymmetries. This anisotropy likely explains why the spherically averaged third-order structure function does not recover the analytically derived $-4/3$ constant, as observed in \citet{gomes2015energy}. We also find that the interscale energy transfer due to the mean flow is negligible throughout scale space, in contrast to results from the von Kármán flow by \citet{knutsen_inter-scale_2020}, which show local magnitudes of the order of the nonlinear transfer. This supports the effectiveness of increasing the number of forcing directions to achieve near-ideal turbulence conditions. Overall, these results evidence the presence of a forward energy cascade where the energy injected by the external forcing is transferred, on average, at a rate $\varepsilon$ down to smaller scales before being dissipated.

From a temporal perspective, we analyze the flow variability by considering the Eulerian frequency spectra. We find that our results agree well with the random sweeping hypothesis adopted by \citet{tennekes1975eulerian}. This implies that, instead of an inertial subrange where statistics are uniquely determined by the dissipation rate, an inertial-convective subrange emerges, where statistics are governed by both the integral velocity scale and the dissipation rate. We show that the Eulerian frequency spectra obtained for different propeller frequencies collapse when normalized using the integral velocity scale and the dissipation rate, the latter having been validated in the previous section through its equivalence with the interscale transfer rate and its consistency with Kolmogorov’s two-thirds law. We also report the constant of the inertial-convective part of the energy spectrum proposed by \citet{tennekes1975eulerian} and compare it with values available in the literature, corresponding mostly to anisotropic flows. As pointed out by \citet{Variano2008}, studying the Eulerian frequency spectra under varying conditions is crucial for improving the estimation of dissipation rates from point velocity measurements in flows without a dominant mean velocity, a condition where the Taylor's frozen turbulence hypothesis does not hold.

Finally, we assess from a global perspective how external forcing is connected to the turbulent scales characterizing the energy cascade. Following dimensional arguments conventionally applied to stirred tanks and von Kármán devices, we show that the dissipation rate, which in the inertial subrange describes on average the complex dynamics discussed in Sec.~\ref{sec:khh}, can be related to the input power in propeller-driven HIT chambers. Similarly, the integral velocity scale, which along with $\varepsilon$ determines the statistics in the inertial-convective range observed in Sec.~\ref{sec:eulerian_spectra}, can be linked to the external variables by introducing a dimensionless quantity that characterizes the axial flow generated by the propellers. This framework correctly predicts the scaling of both the dissipation rate and the integral velocity scale with propeller frequency, and consequently that of derived quantities such as the Taylor and Kolmogorov scales. These predictions are validated by both our measurements and data from previously reported HIT chambers, covering a wide range of chamber and propeller designs, number of propellers, and working fluids. We find that the two constants associated with the dissipation and velocity scales remain within one order of magnitude across nearly all devices, despite the diversity of experimental configurations. The integral length scale, in contrast, is obtained by considering dynamic equilibrium at large scales, which introduces the dissipation coefficient $C_\varepsilon$. While most studies do not allow its independent determination, from our data we obtain a value of $C_\varepsilon \approx 0.5$, in close agreement with results of DNS of forced HIT \citep{mccomb2015energy}. However, a comparison with the results of \citet{wang2025lagrangian} suggests that the specific propeller design may influence the value of this coefficient. We hope this approach will motivate the direct measurement of external variables, such as the torque or power consumption of the motors, in propeller-driven HIT chambers, as is already common in the analysis of von Kármán devices \citep{cadot1997energy,mordant1997characterization,burnishev_torque_2014} and mixing vessels \citep{sirasitthichoke2017power,steiros2017effect}.

\begin{acknowledgments}
This work has been supported by ANID/CONICYT FONDECYT Regular 1210936 and ANID/CONICYT FONDEQUIP EQM180150. L.H. acknowledges the scholarship ANIDPFCHA/Doctorado Nacional/2019-21191759.
\end{acknowledgments}

\bibliography{bibfile}

\end{document}